\documentclass[aps,prl,preprintnumbers,amsmath,amssymb,superscriptaddress,twocolumn]{revtex4-2}
\usepackage{amsmath}
\usepackage{hyperref}
\hypersetup{colorlinks=true, citecolor=blue, urlcolor=blue, linkcolor=blue}
\usepackage{amssymb}
\usepackage{hyperref}
\usepackage{graphicx} 
\usepackage[capitalize]{cleveref}
\usepackage{lipsum}
\usepackage{comment}
\usepackage{xcolor}

\begin{document}

\newcommand{\td}{\,\mbox{\rm d}}
\newcommand{\be}{\boldsymbol{e}}
\newcommand{\bs}[1]{\vec{\boldsymbol{#1}}}
\newcommand{\revzhx}[1]{\textcolor{red}{#1}}

\title{Designing Coherent Optical Environment for Dynamic Optical Manipulation with a Simple Control Beam}

\author{Xiaoshu Zhao}
\affiliation{State Key Laboratory of Surface Physics and Department of Physics, Fudan University, Shanghai 200433, China}

\author{Xu Yuan}
\affiliation{School of Electronic Engineering, Guangxi University of Science and Technology, Liuzhou, Guangxi 545006, China}

\author{Hongxia Zheng}
\email[]{hxzheng18@fudan.edu.cn}
\affiliation{School of Electronic Engineering, Guangxi University of Science and Technology, Liuzhou, Guangxi 545006, China}
\affiliation{Guangxi Key Laboratory of Multidimensional Information Fusion for Intelligent Vehicles, Liuzhou, Guangxi 545006, China}
\affiliation{State Key Laboratory of Surface Physics and Department of Physics, Fudan University, Shanghai 200433, China}

\author{Huajin Chen}
\email[]{huajinchen13@fudan.edu.cn}
\affiliation{School of Electronic Engineering, Guangxi University of Science and Technology, Liuzhou, Guangxi 545006, China}
\affiliation{Guangxi Key Laboratory of Multidimensional Information Fusion for Intelligent Vehicles, Liuzhou, Guangxi 545006, China}
\affiliation{State Key Laboratory of Surface Physics and Department of Physics, Fudan University, Shanghai 200433, China}

\author{Zhifang Lin}
\email[]{phlin@fudan.edu.cn}
\affiliation{State Key Laboratory of Surface Physics and Department of Physics, Fudan University, Shanghai 200433, China}
\affiliation{Collaborative Innovation Center of Advanced Microstructures, Nanjing University, Nanjing 210093, China}

\begin{abstract} 
We propose a framework for designing coherent optical environments that enable versatile and dynamic optical manipulation. 
In contrast to conventional material-based near-field platforms, our approach employs a structured coherent light field---optimized via a back-propagation-based inverse design algorithm---as the manipulation environment. This light-based platform allows a simple control beam, 
such as a single plane wave or a low-numerical-aperture Gaussian beam, to steer micro-objects effectively. 
By establishing a one-to-one correspondence between 
control beam parameters (e.g., 
phase/polarization of a plane wave) and 
particle trapping positions, our method enables real-time and versatile control of particles. 
A wide range of two- and three-dimensional trajectories---including circles, squares, tree-like paths, and epicycle-deferent curves---can be achieved solely by modulating the phase of the control beam. 
This design strategy for the structured-light environments offers a dynamically reconfigurable, 
all-optical, and contact-free platform for advanced optical manipulation in free space, 
with promising applications in nanorobotics, biological probing, and beyond.
\end{abstract}

\maketitle

Optical manipulation has emerged as a powerful and versatile technique for exerting precise control over microscopic and nanoscopic objects through optical forces and torques. It has found widespread applications across disciplines including physics, chemistry, biology, materials science, nanotechnology, 
and biomedical engineering \cite{ashkin1970,ashkin1986,Ashkin1997,Guck2001,Grier2003,Neuman2004,Juan2011,Marago2013}. 
The field of optical manipulation has traditionally advanced through two complementary strategies: engineering either the manipulating light field or the target object.
Structuring the spatial amplitude and phase profile of light---such as in Airy beams  \cite{airybeam2007}, 
which guide particles along curved trajectories  \cite{airybeam2008}, and Bessel beams \cite{Durnin1987},
which realize optical pulling \cite{pullingchen2011,WOS:000297132200002} against the wave propagation direction---has 
enabled a broad range of advanced manipulation functionalities
\cite{holographictweezers2002,Advphotonics2021,Vortex2010,airybeam2008,pullingchen2011,WOS:000297132200002,ImPoynting2022}. 
Concurrently, tailoring the morphology and/or optical response of particles has 
revealed 
additional manipulation phenomena
\cite{WOS:000335011700015,SciAdvpulling2019,opticallift2011,PhaseChang2025,Janusparitcle,WOS:000712284300001,WOS:001323979500001}, 
enabling effects like lateral optical forces that 
induce optical lift in an anisotropic particle  \cite{opticallift2011}
and reversible trapping-release transitions with phase-change materials  \cite{PhaseChang2025}. When synergized, 
these two approaches have elevated optical manipulation from static trapping to dynamic and programmable control, 
enabling applications such as nanoscale assembly, biological force probing, and beyond
\cite{molecularknot1999,WOS:000187719300038,WOS:000508458700001,Marago2013,WOS:000552283000001}.

Recently, a paradigm shift has emerged in optical manipulation—toward 
engineering the near-field environment
\cite{WOS:001269001300006,WOS:001319882200001,WOS:000516663000001,WOS:000645436800017,WOS:000630005400001,WOS:001456600400001,WOS:001110458000005,NatPhotLinearmom2013,WOS:001297264800001,WOS:000536569900002,WOS:000537235300017}.
By embedding particles within structured photonic media---including metasurfaces, 
metamaterials, photonic crystals, and topological materials---researchers have opened a new avenue for generating a range of distinctive optical forces,
such as lateral forces, pulling forces, and
enantioselective forces \cite{WOS:000841997700001,WOS:000841893000001,WOS:000271374600068,WOS:000731589200001,WOS:001456028800001,WOS:000954618400001,topology2020}. 
These effects arise from complex near-field interactions mediated by the photonic environment engineered through material-based structures. 
However, the intrinsic rigidity, limited reconfigurability, and fabrication complexity of such material-based systems 
constrain their adaptability in dynamic or reprogrammable applications. 

In this Letter, we propose a dynamically reconfigurable, all-optical alternative:
replacing the static material-based environments with an intricately sculpted coherent optical environment,
which serves as the manipulation background. 
The desired manipulation functionalities is embedded into 
the coherent structured-light environment \cite{structuredlightJOpt2017,structuredlight2019,structuredlight2021} optimized via the back-propagation algorithm \cite{backpropagation1986,Werbos1994,nnrev2015}.
The optimization enables even a simple low-NA Gaussian beam or a plane wave, termed the control beam, 
to exert sophisticated control over particles via coherent coupling with the engineered optical environment. 
Our numerical simulations demonstrate that the resulting framework 
functions as a ``phase-to-position transducer'': continuous modulation of the control beam’s phase results in smooth 
and controllable particle displacement along predefined trajectories. 
This mechanism allows robust, real-time, and in situ reconfiguration through phase-only modulation, 
requiring no reshaping control beams or structured-light environments, or reconfiguration of microstructures
Furthermore, the absence of material 
scaffolds 
enables unobstructed particle motion, thereby enhancing the flexibility of optical manipulation.
These results promise a versatile platform for optical control, with potential applications in vivo biological probing and dynamic nanorobotics.

\begin{figure*}
	\centering
	\includegraphics[width=0.9 \textwidth]{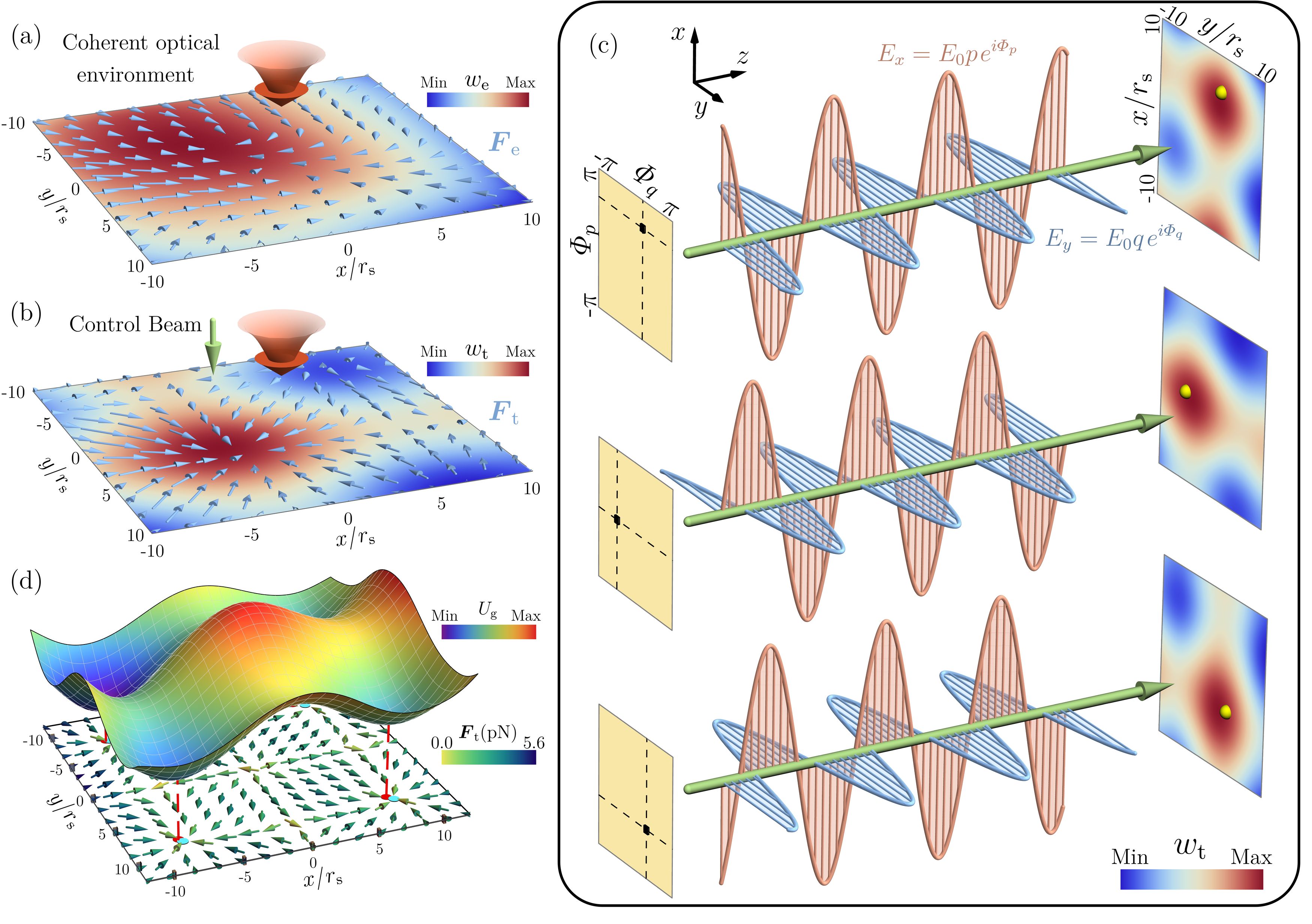}
\caption{\textbf{(a)} The electromagnetic energy density \( w_\text{e} \) distribution (color map) 
and the corresponding in-plane optical force \( \boldsymbol{F}_\text{e} \) (overlaid vectors) at $z=0$ plane
generated by the optimized coherent optical environment (a designed structured-light) 
schematically represented by the orange conical-shaped arrow. 
This environment is designed 
via back-propagation algorithm to interfere 
with a single plane wave (control beam) for trapping the particle
at arbitrary positions within a square region by phase modulation of the control beam.
\textbf{(b)} Total electromagnetic energy density \( w_\text{t} \) distribution (color map) and the corresponding in-plane optical force
\( \boldsymbol{F}_\text{t} \) (overlaid vectors) produced by the interference 
of the coherent optical environment 
shown in (a) and a plane-wave control beam (shown schematically by the green arrow) 
with polarization phase \( \varPhi_p = -\varPhi_q = -\pi/3 \), see, Eq.~\eqref{eq2}.
\textbf{(c)} Three representative cases demonstrating the dynamic ``phase-to-position transducer'' effect, where the
in-plane trapping position \((x, y)\) is determined by the plane-wave control beam. 
By dynamically adjusting the phase, the particle can be smoothly repositioned along arbitrary trajectories
within the region of $8r_{\text s}\times 8r_{\text s}$ centered at the origin $(x,y)=(0,0)$.
The displayed results correspond to $(\varPhi_p, \varPhi_q)=(0.6\pi,0.2\pi), (0.0\pi,-0.5\pi), (-0.3\pi,0.3\pi)$ (top to bottom).
Also shown are $w_{\text t}$ distribution (color map) and schematic plots of the control beam.
\textbf{(d)} Optical potential $U_\text{g}$ and 
total optical force $\boldsymbol{F}_\text{t}$ exerted on the particle  
for $\varPhi_p=\varPhi_q=\pi$.  The potential minima (red dots) and 
trapped sites from molecular statics analysis (cyan dots) are shown. 
The slight discrepancy arises because the optical force consists of both the conservative and non-conservative components 
\cite{Sukhov2017,jiang2016decomposition,WOS:000486622300012,zheng2020decompchiral}. 
}
	\label{fig:2D_main2}
\end{figure*}

To illustrate the fundamental concept 
of the coherent optical environment, consider 
an isotropic particle suspended in water with
permittivity $\varepsilon_{\text s}=2.53$ and radius $r_{\text s}=\lambda_\text{b}$, where $\lambda_\text{b}=0.8 \, \mu \text{m}$ is the
wavelength of the control beam in water.
We aim to control the motion of the Mie-sized particle in a ``phase-to-position transducer'' manner:
by tuning the phase of a plane-wave control beam, the particle can be trapped at a predetermined position, 
with each phase uniquely 
mapping to a specific trap site. Continuous phase modulation then allows for smooth relocation of
the trapped particle along a desired  
trajectory. 
To this end, we first write the total electric field ($E$-field) incident on the particle as 
\begin{equation}\label{eq1}
\boldsymbol{E} = \boldsymbol{E}_{\text c}+\boldsymbol{E}_\text{e},
\end{equation}
where $\boldsymbol{E}_{\text c}$ is the plane-wave control beam propagating along $\hat{\boldsymbol z}$ (or $-\hat{\boldsymbol z}$)
and $\boldsymbol{E}_\text{e}$ is the structured-light environment 
(the time dependence $e^{-i\omega t}$ is assumed):
\begin{equation}
\boldsymbol{E}_{\text c} = 
    E_0(p\,e^{i\varPhi_p}\hat{\boldsymbol x}+q\,e^{i\varPhi_q}\hat{\boldsymbol y})e^{\pm i kz}, \label{eq2} 
\end{equation}
\begin{equation}
\boldsymbol{E}_\text{e} = 
    \sum_{j=1}^{n_{\text p}}\boldsymbol{\mathcal{E}}_j \,e^{i k\,\hat{\boldsymbol{k}}_j\cdot \boldsymbol{r} }
    \; \; \text{with }
    \boldsymbol{\mathcal{E}}_j= E_0 (p_j\hat{\boldsymbol{\theta}}_j + q_j\hat{\boldsymbol{\varphi}}_j).
\label{eq3}
\end{equation}
Here we represent $\boldsymbol{E}_\text{e}$ as a superposition of $n_{\text p}$ plane waves to facilitate optimization via back-propagation. 
This approach is justified for any a spatial structured-light field, see, e.g., \cite{zhao2023nanophoton},
and aligns with experimental implementation using the spatial light modulators \cite{weiner2000femtosecond,zhu2014arbitrary,forbes2016SLM}, 
bridging theoretical design and practical realization. 
In Eq.~\eqref{eq3},
$\boldsymbol{\mathcal{E}}_j$ represents the complex amplitude vectors for the $j$-th constituent plane wave 
comprising the structured-light environment,
with the polarization governed by $p_j$ and $q_j$. 
The unit vectors $\hat{\boldsymbol{\theta}}_j$ and $\hat{\boldsymbol{\varphi}}_j$ correspond to
increasing polar and azimuthal angles, respectively, and, together with the wavevector direction $\hat{\boldsymbol{k}}_j=\boldsymbol{k}_j/k$,
form a right-handed triplet ($\hat{\boldsymbol{\theta}}_j\times\hat{\boldsymbol{\varphi}}_j=\hat{\boldsymbol{k}}_j$), with
$k={2\pi}/{\lambda_\text{b}}$ denoting the wavenumber in water. The parameter
$E_0$ sets the energy intensity of the control beam. 
In all our simulation, $E_0$ is set to a value such that the control beam irradiance 
$I_0=|\boldsymbol{E}_{\text{c}}|^2/(2Z_\text{b})=1.0$\,mW$/\mu$m$^2$, where
$Z_\text{b}$ is the wave impedance of water. Finally, $n_{\text p}$ serves as a hyperparameter in optimization.
The structured-light environment 
is optimized so that adjusting
only the control beam's phase $(\varPhi_p,\varPhi_q)$ can relocate the trapped particle to predefined position, 
with each phase $(\varPhi_p,\varPhi_q)$ uniquely 
determining a trap site.

In addition to the parameters $P=\big\{p_j, q_j, \hat{\boldsymbol{k}}_j\big|j=1,2,\dots,n_p\big\}$ that 
define the plane waves composing the structured-light environment, the complex polarization coefficients
$p$ and $q$ of the control beam must also be optimized to achieve the desired ``phase-to-position transducer'' behavior.
The optimization proceeds as follows. Beginning with randomized values for the full parameter set $P_{\text{opt}}=P\bigcup\{p,\,q\}$, we
compute the optical force using the rigorous analytical framework of Cartesian multipole expansion theory 
\cite{jiang2015universal,jiang2016decomposition,WOS:000486622300012,zheng2020decompchiral}. This approach efficiently computes 
the optical forces on particle in coherent multi-plane wave fields \cite{zhao2023nanophoton}, circumventing the need for special functions (e.g., the associated Legendre functions) that interfere with the auto-gradient computation in the back-propagation-based optimization. 
With the Cartesian multipole expansion, 
the time-averaged optical force $\boldsymbol{F}$ takes a compact form:
\begin{equation}\label{eq:Fij}
\boldsymbol{F}
=\text{Im} \sum_{i,j} \boldsymbol F_{ij} \, e^{i\,(\boldsymbol{k}_i-\boldsymbol{k}_j) \cdot \boldsymbol{r}},
\end{equation}
where $\boldsymbol{F}_{ij}$ is $\boldsymbol{r}$ (particle position) independent, and the sum is taken over
the control beam and $n_{\text p}$ plane waves constituting the structured-light environment.
Next, the optimization loop iterates through the following steps: 1) evaluating a tailored loss function with some regularization terms; 
2) computing its gradient with respect to each parameter in $P_{\text{opt}}$ via back-propagation; and,
3) updating $P_{\text{opt}}$ using the Adam optimizer \cite{kingma2017adam} to minimize the loss.
The explicit form of  $\boldsymbol F_{ij}$ and 
further optimization details are given in the supplementary material.
Our implementation leverages Pytorch \cite{torch2019} for automatic differentiation and SciPy library \cite{2020SciPy-NMeth}
for computing Legendre polynomials and their derivatives.

Figure~\ref{fig:2D_main2} 
demonstrates our dynamic in-plane optical trapping scheme
for a Mie-sized particle within a $8r_{\text s}\times 8r_{\text s}$ square region, 
enabled by the superposition of an optimized coherent structured-light environment, see, Eq.~\eqref{eq3}, and
a phase tunable plane-wave control beam propagating in $\hat{\boldsymbol{z}}$ defined by Eq.~\eqref{eq2} with the upper sign. 
The parameter set $P_{\text{opt}}$ is fully optimized to achieve the ``phase-to-position transducer'': by modulating only the control beam's phase $(\varPhi_p,\varPhi_q)$, the particle can be relocated to any target position within the square region.

Figure~\ref{fig:2D_main2}(a) shows the intensity profile $w_\text{e}$ of the optimized coherent optical environment, which initially traps the particle near the square region's center, as indicated by the force profile $\boldsymbol{F}_\text{e}$. Superimposing the control beam can thus reposition
the particle to a new site 
determined uniquely 
by the phase  $(\varPhi_p,\varPhi_q)$.
For example, Fig.~\ref{fig:2D_main2}(b) displays 
the total electromagnetic intensity $w_{\text{t}}$ and the total in-plane optical force $\boldsymbol{F}_{\text t}$ 
for $(\varPhi_p,\varPhi_q)=(-\pi/3,\pi/3)$.
The ``phase-to-position'' mapping is further illustrated in
Fig.~\ref{fig:2D_main2}(c), where
three representative phases dynamically relocate the particle to different sites. 
This mapping cover the entire phase space $-\pi <\varPhi_p, \varPhi_q< \pi$ 
and target region, with a nearly uniform trapping stiffness
of around \(1.45\, \mathrm{pN}/\mu\mathrm{m}\). Consequently, 
the particle can be smoothly guided along arbitrary trajectories or precisely positioned at any desired cites, enabling real-time manipulation simply through phase tuning of a single plane-wave control beam. 

Notably, four phase combinations \((\varPhi_p, \varPhi_q) = (\pm\pi, \pm\pi)\) generate identical optical fields, each producing the same set of four trapping sites arranged at the corners of a square, as shown in Fig.~\ref{fig:2D_main2}(d) for the representative case $(\pi,\pi)$.
The depth of optical potential $U_{\text g}$ varies across the four trapping sites, since it is obtained by
integrating the conservative component of the optical force. Nevertheless,
the trapping stiffness remains consistent as it is obtained by total optical force 
consisting of both conservative and non-conservative components. 
For the Mie-sized particles, the non-conservative optical force becomes significant. Although 
the conservative force still dominates in this case, neglecting the
non-conservative component introduces trapped position deviations of
\(0.25 \,r_\text{s}\) to \(r_\text{s}\), see, the red and cyan dots in Fig.~\ref{fig:2D_main2}(d). 
Our inverse design is a scheme that optimizes the structured-light environment using directly the total optical force 
(conservative plus non-conservative), unlike intensity-based approaches, which are actually based 
exclusively on the conservative force. 
By accounting for all physical contribution to the trapping, the total force-based framework enables 
more precise particle manipulation.

\begin{figure}[ht]
    \centering
    \includegraphics[width=\columnwidth]{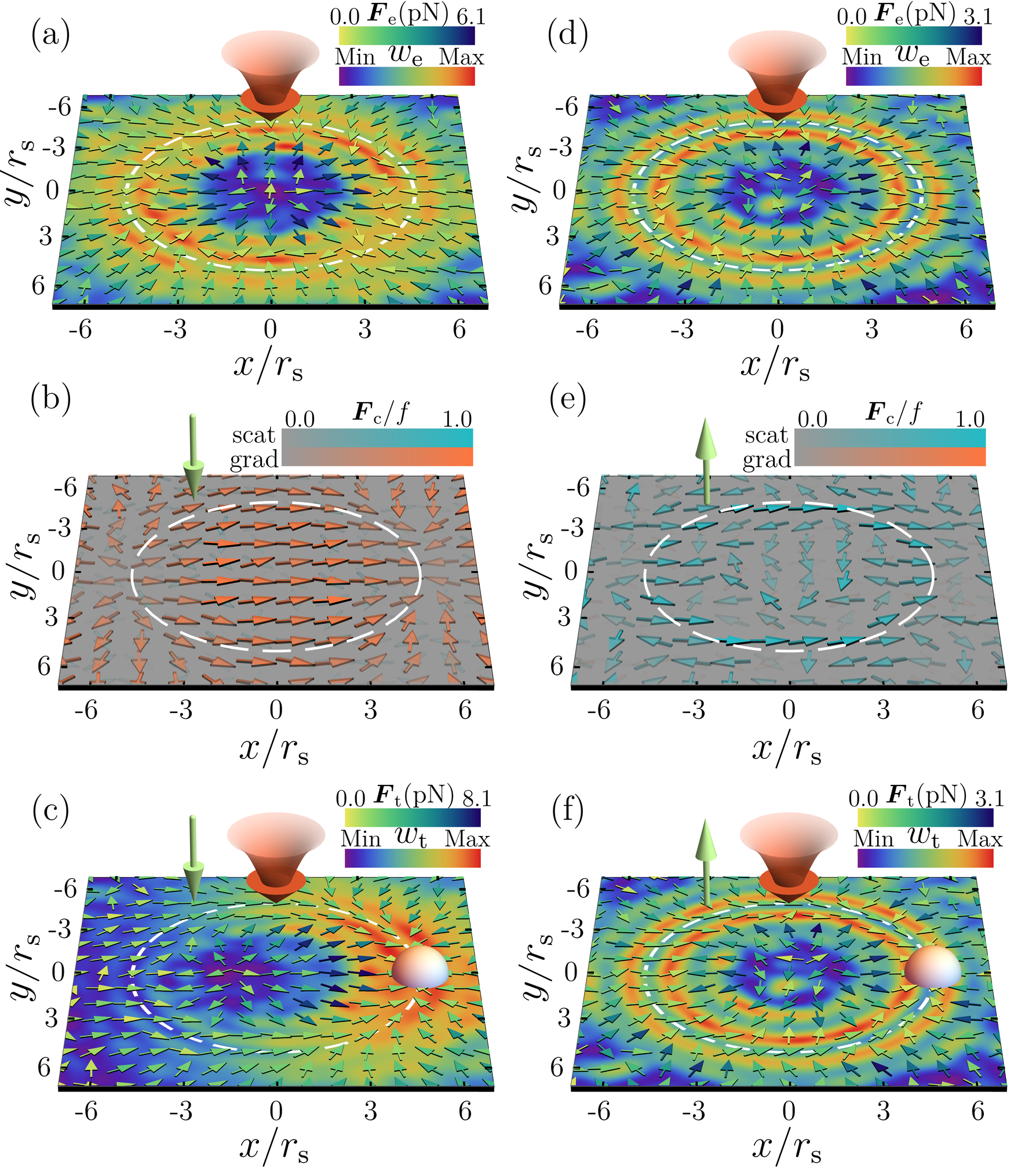}
    \caption{Two schemes for particle manipulation along a predefined ring-shaped trajectory (white dashed line) 
    using control beams (schematically represented by green arrows) propagating in $\hat{\boldsymbol{z}}$, Case I (a-c), and in
    $-\hat{\boldsymbol{z}}$, Case II (d-f). In both cases, the coherent structured-light environments (orange conical arrows) consist of
    plane waves with positive $k_z$ components. 
    \textbf{(a, d)}: 
        The energy density profiles $w_\text{e}$ (color map) of the optimized environment 
        for Cases I and II, respectively. Overlaid vectors represent in-plane optical forces $\boldsymbol{F}_\text{e}$  
        generated by the optical environment, demonstrating 
        particle confinements near the target trajectories.
    \textbf{(b, e)}:
        The ``control-induced'' optical forces 
        \(\boldsymbol{F}_\text{c} \equiv \boldsymbol{F}_\text{t} - \boldsymbol{F}_\text{e}\),
        where $\boldsymbol{F}_\text{t}$ is the total force from the combined field of the environment and the control beam. 
        The forces $\boldsymbol{F}_\text{c}$  
        are decomposed into gradient ($\boldsymbol{F}_\text{c}^{\text{grad}}$) and 
        scattering ($\boldsymbol{F}_\text{c}^{\text{scat}}$) components, normalized by
        \(f = \max\left( \left| \boldsymbol{F}_\text{c}^\text{grad} \right|, \left| \boldsymbol{F}_\text{c}^\text{scat} \right| \right)\). 
    \textbf{(c, f)}:
        The energy density $w_\text{t}$ 
        and in-plane optical force $\boldsymbol{F}_\text{t}$ generated by the total field.
        Both schemes achieve stable in-plane trapping at $(x,y)=(5 r_{\text s},0)$ for $\varPhi=0$, but 
            rely on the gradient (Case I) and scattering (Case II) forces, respectively.
    }
    \label{fig:circle_main}
\end{figure}

While the dual-phase $(\varPhi_p,\varPhi_q)$ modulation scheme provides full positional control within the $8 r_{\text s}\times8r_{\text s}$ 
region, we next demonstrate a simplified trajectory-following paradigm requiring only a single phase parameter $\varPhi$ by setting 
$\varPhi_p=\varPhi_q=\varPhi$. This establishes a one-to-one mapping between phase $\varPhi$ 
and particle position along a predefined path, preserving the ``phase-to-position transducer'' effect while reducing control dimensionality. 
Continuous phase modulation then enables direct particle guidance along arbitrary predefined trajectories.
In this scheme, 
the optimized structured-light environment 
ensures confinement of the particle near the target trajectory, while exhibiting negligible positional bias along the path,
as evidenced by the force profile $\boldsymbol{F}_\text{e}$ in Figs.~\ref{fig:circle_main}(a) and (d). 
Superimposing a plane-wave control beam allows precise positioning anywhere along path through phase $\varPhi$ modulation alone.
Figures~\ref{fig:circle_main}(a–c) (Case I) and Figs.~\ref{fig:circle_main}(d–f) (Case II) 
illustrate two implementations of this scheme, 
both designed to position the particle on a ring-shaped path of radius~$5r_{\text{s}}$ 
via phase tuning.
In both configurations, the structured-light environment comprises plane waves with positive~$k_z$ components, while the control beam propagates either co-directionally (Case~I, along~$\hat{\boldsymbol{z}}$) or counter-directionally (Case~II, along~$-\hat{\boldsymbol{z}}$) with respect to the environment.
The structured-light environment for each configuration is optimized separately, thus showing distinct energy density 
$w_\text{e}$ profiles (Figs.~\ref{fig:circle_main}(a) and (d)). 
In both cases, $\boldsymbol{F}_\text{e}$ and $w_\text{e}$ profiles confirm that the structured-light environment alone confines the particle via gradient forces. In contrast, the nature of the ``control-induced'' force, defined as $\boldsymbol{F}_\text{c} \equiv \boldsymbol{F}_\text{t} - \boldsymbol{F}_\text{e}$, differs markedly, as illustrated for $\varPhi = 0$ in Figs.~\ref{fig:circle_main}(b) and (e).
In Case I,
$\boldsymbol{F}_\text{c}$ is dominated by the conservative forces, 
with non-conservative components negligible (Fig.~\ref{fig:circle_main}(b)). 
It is noted that we have plotted vectors for both the gradient and 
scattering forces (or, equivalently, the conservative and non-conservative forces). The
scattering force vectors are nearly graphically indiscernible due to their negligible magnitude.
In Fig.~\ref{fig:circle_main}(e) for Case II, 
however, $\boldsymbol{F}_\text{c}$ is governed by non-conservative forces.
Despite these fundamental differences, both cases achieve stable in-plane trapping at 
the target position $(x,y)=(5r_{\text s},0)$ for $\varPhi=0$,
with stiffnesses of 0.5 \(\text{pN}/\mu\text{m}\) (Case I) and 0.2 \(\text{pN}/\mu\text{m}\) (Case II), 
as demonstrated by the total force profiles $\boldsymbol{F}_\text{t}$ in Figs.~\ref{fig:circle_main}(c) and (f).

\begin{figure}[t]
    \centering
    \includegraphics[width=\columnwidth]{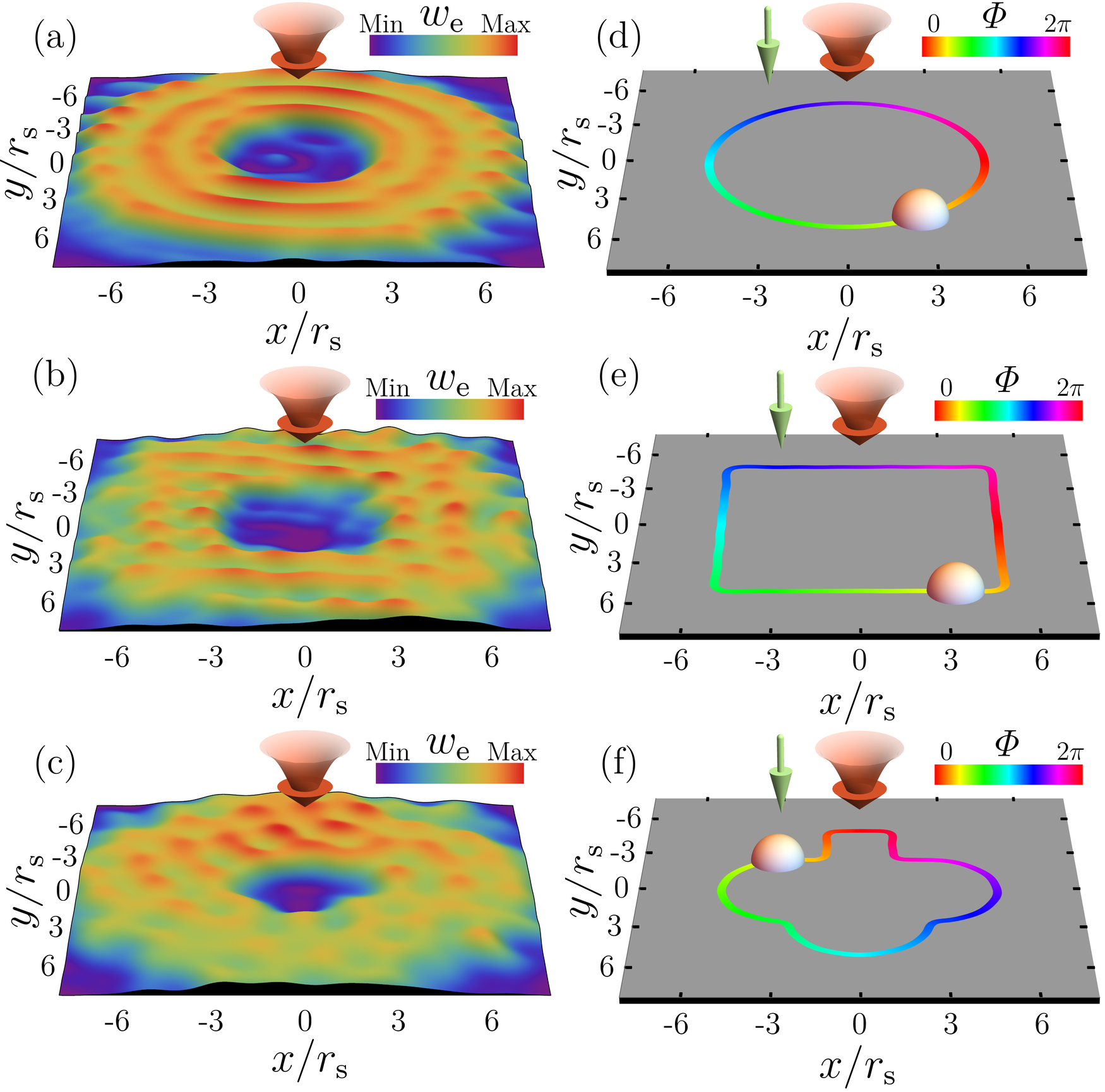}
    \caption{Phase-controlled optical manipulation along three distinct trajectories 
    using the co-propagating configuration with linear polarization ($E_y=0$) for both control beams and 
    structured-light environments.   
\textbf{(a, d)}:
     Ring-shaped trajectory: (a) Optimized structured-light environment intensity $w_\text{e}$ distribution; (d) Phase-position mapping.
\textbf{(b, e)}:
     Smoothed-corner square trajectory: (b) 
     $w_\text{e}$ profile; (e) Phase-controlled positioning. 
\textbf{(c, f)}:
     Tree-shaped trajectory: (c) 
     $w_\text{e}$ distribution; (f) Phase-based positioning result.
     All particle positions are computed via molecular statics analysis.
     The spheres in (d-f) show trapping sites for $\varPhi=\pi/3$.}
    \label{fig:line_all}
\end{figure}

The trajectory-following paradigm is not confined to ring-shaped paths. 
We now demonstrate its implementation on more complex trajectories while
simplifying the configuration.
Since the counter-propagating scheme relies on the non-conservative force for trapping,   
limiting its practicality \cite{opticalearnshaw1983}, 
we focus on the co-propagating configuration. 
To further facilitate experimental realization, we simplify the polarization configuration by
setting $E_y=0$ for all plane waves composing the structured-light environment, as well as the plane-wave control beam itself. This linear polarization scheme,
more compatible with the standard spatial light modulators \cite{weiner2000femtosecond,zhu2014arbitrary,forbes2016SLM},
reduces the electric fields $\boldsymbol{E}_{\text c}$ and $\boldsymbol{E}_\text{e}$ from Eqs.~\eqref{eq2} and \eqref{eq3} to
\begin{eqnarray} 
\boldsymbol{E}_{\text c}  &\!=&\!
    E_0\, p\,e^{i\varPhi}\hat{\boldsymbol x}\,e^{i kz}, \label{eq2a} \label{Ecx}\\
\boldsymbol{E}_\text{e}  &\!=&\!
    \sum_{j=1}^{n_{\text p}}E_0\, p_j(\hat{\boldsymbol x} \cos\theta_j-\hat{\boldsymbol z}\sin\theta_j \cos\varphi_j) 
                \,e^{i \boldsymbol{k}_j\cdot \boldsymbol{r} },
\label{eq3a}
\end{eqnarray}
where $\theta_j$ and $\varphi_j$ are, respectively, the polar and azimuthal angles of wavevector $\boldsymbol{k}_j$. 
The optimized parameter set now becomes $P_{\text{opt}}={p}\bigcup\big\{p_j, \hat{\boldsymbol{k}}_j\big|j=1,\dots,n_{\text p}\big\}$.

Figure~\ref{fig:line_all} demonstrates the phase-to-position control on three distinct trajectories, which are ring-shaped path (Fig.~\ref{fig:line_all}(a, d)),
smoothed-corner square (Fig.~\ref{fig:line_all}(b, e)), and
tree-shaped configuration (Fig.~\ref{fig:line_all}(c, f)).
The structured-light environment intensity profiles $w_\text{e}$ (Fig.~\ref{fig:line_all}(a-c)) and corresponding particle position
(Fig.~\ref{fig:line_all}(d-f)) reveal precise trapping along each trajectory,
with position uniquely determined by the control beam's phase $\varPhi$ in Eq.~(\ref{Ecx}). The trapping sites are obtained by total optical force-based 
molecular statics analysis, which also yields trapping 
stiffness ranging from  0.15 to $0.6~ \mathrm{pN}/\mu\mathrm{m}$ across all configurations shown.
Continuous phase modulation enables smooth particle guidance along these paths, suggesting potential applications as an optical synchronous motor.

\begin{figure}
    \centering
    \includegraphics[width=0.85 \columnwidth]{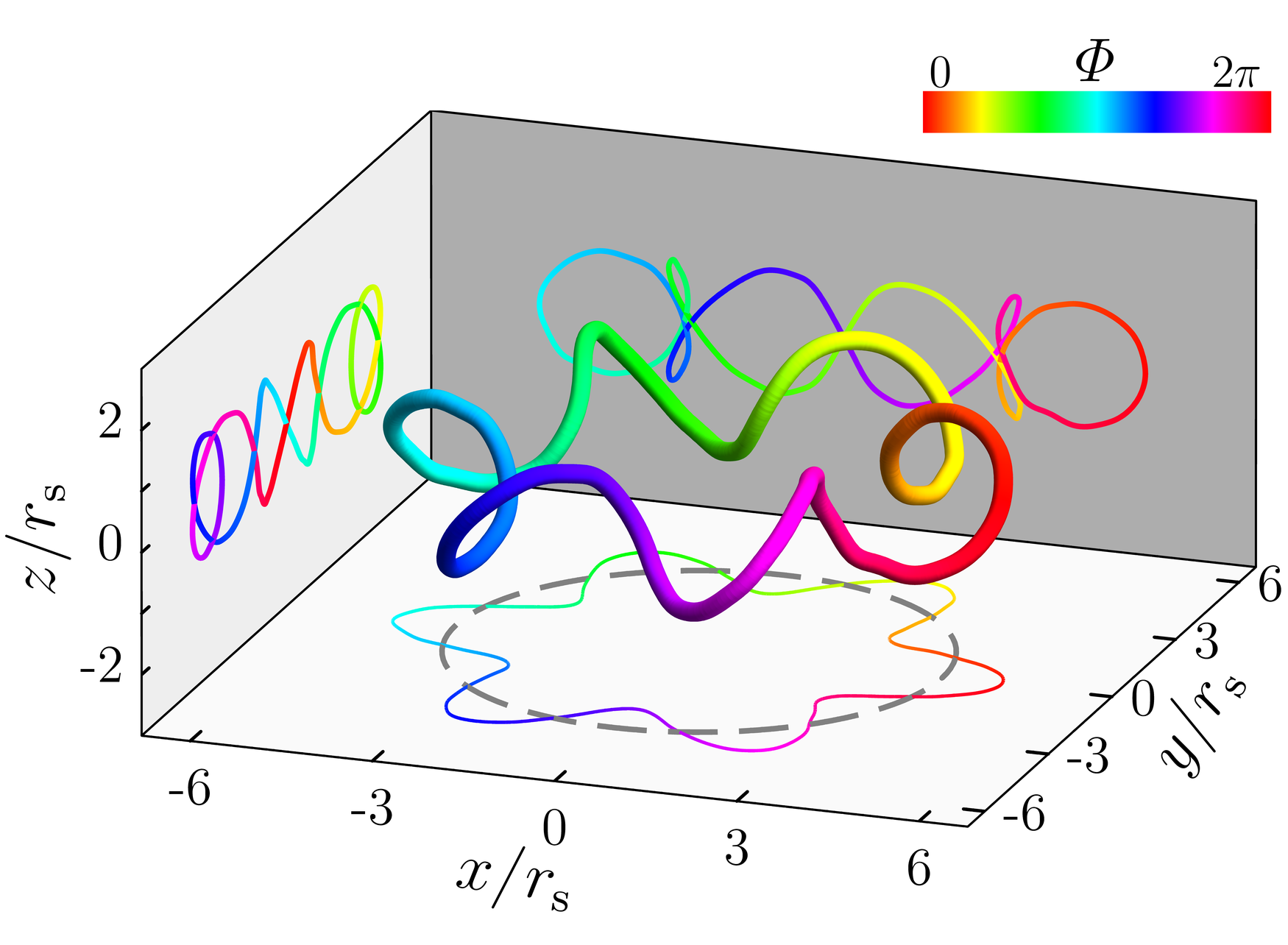}
    \caption{Three-dimensional (3D) phase-to-position control of particle along an epicycle-deferent trajectory. 
    The 3D rendering shows particle positions (color-coded by $\varPhi$) and three orthogonal plane projections,
    demonstrating full 3D position control by modulating control beam's phase $\varPhi$.
    All positions are computed via molecular statics analysis of total optical force from combined field.
    }
    \label{fig:3D_main}
\end{figure}

The phase-to-position mapping paradigm extends naturally to three-dimensional (3D) dynamic manipulation using a single plane-wave control beam. 
Figure~\ref{fig:3D_main} demonstrates such a mapping (or guided particle transport) along an
epicycle-deferent trajectory, confirming 3D control via single-phase modulation. It is achieved by a plane-wave control beam 
imposed on an intricately designed structured-light environment. The latter is 
comprised of a co-propagating component (coherent with the control beam) and a counter-propagating incoherent component.
This 3D path consists of a primary deferent circle (radius \(4r_\text{s}\)) centered at the origin, and,
a perpendicular epicycle (radius \(r_\text{s}\)) completing six revolutions per deferent cycle (6:1 ratio).
The resulting particle position \(\boldsymbol{r}\) is parameterized 
in Cartesian coordinates by the dynamic phase \(\varPhi\) as
\begin{equation}\label{eq:3D_position}
	\boldsymbol{r} = \big(d \, \cos \varPhi,\, d \, \sin \varPhi,\, r_\text{s} \,\sin (6 \, \varPhi) \big),
\end{equation}
where \(d = 4\,r_\text{s} + r_\text{s} \cos (6\,\varPhi)\). The confinement of the particle  
remains stable throughout the intricate trajectory, with trapping stiffness ranging from 
\(0.05\) to \(0.2\,\mathrm{pN}/\mu\mathrm{m}\).

In summary, we have demonstrated a paradigm for dynamic optical manipulation by replacing the traditional material-based environments 
with optimized coherent optical environments. 
Through inverse design using back-propagation-enabled optimization 
based on Cartesian multipole expansion theory for the optical force calculation, 
these optical environments synergize with a simple phase-modulated plane wave to enable real-time particle control. 
Our theoretical framework demonstrates highly versatile manipulation capabilities—ranging from arbitrary positioning and path guidance within a 2D region, to controlled motion along complex 2D (e.g., circular, square, tree-like) and 3D (e.g., epicycle–deferent) trajectories. Remarkably, all functionalities are achieved solely by modulating the phase of a single plane-wave control beam.
This approach establishes structured-light environments as a powerful platform for optical force engineering, 
with transformative potential for applications in nanoscale assembly, active matter control, and biophotonics, 
where unobstructed, scalable manipulation is crucial.

\begin{acknowledgments}
This work is supported by National Natural Science Foundation of China (No. 12074084, No. 12204117, and No. 12174076); 
Guangxi Science and Technology Projects (2024GXNSFBA010261, AD23026117,  and 2023GXNSFFA026002); 
Open Project of State Key Laboratory of Surface Physics in Fudan University (KF2022\_15).
\end{acknowledgments}

\bibliography{reference}

\begin{thebibliography}{65}%
\makeatletter
\providecommand \@ifxundefined [1]{%
 \@ifx{#1\undefined}
}%
\providecommand \@ifnum [1]{%
 \ifnum #1\expandafter \@firstoftwo
 \else \expandafter \@secondoftwo
 \fi
}%
\providecommand \@ifx [1]{%
 \ifx #1\expandafter \@firstoftwo
 \else \expandafter \@secondoftwo
 \fi
}%
\providecommand \natexlab [1]{#1}%
\providecommand \enquote  [1]{``#1''}%
\providecommand \bibnamefont  [1]{#1}%
\providecommand \bibfnamefont [1]{#1}%
\providecommand \citenamefont [1]{#1}%
\providecommand \href@noop [0]{\@secondoftwo}%
\providecommand \href [0]{\begingroup \@sanitize@url \@href}%
\providecommand \@href[1]{\@@startlink{#1}\@@href}%
\providecommand \@@href[1]{\endgroup#1\@@endlink}%
\providecommand \@sanitize@url [0]{\catcode `\\12\catcode `\$12\catcode `\&12\catcode `\#12\catcode `\^12\catcode `\_12\catcode `\%12\relax}%
\providecommand \@@startlink[1]{}%
\providecommand \@@endlink[0]{}%
\providecommand \url  [0]{\begingroup\@sanitize@url \@url }%
\providecommand \@url [1]{\endgroup\@href {#1}{\urlprefix }}%
\providecommand \urlprefix  [0]{URL }%
\providecommand \Eprint [0]{\href }%
\providecommand \doibase [0]{https://doi.org/}%
\providecommand \selectlanguage [0]{\@gobble}%
\providecommand \bibinfo  [0]{\@secondoftwo}%
\providecommand \bibfield  [0]{\@secondoftwo}%
\providecommand \translation [1]{[#1]}%
\providecommand \BibitemOpen [0]{}%
\providecommand \bibitemStop [0]{}%
\providecommand \bibitemNoStop [0]{.\EOS\space}%
\providecommand \EOS [0]{\spacefactor3000\relax}%
\providecommand \BibitemShut  [1]{\csname bibitem#1\endcsname}%
\let\auto@bib@innerbib\@empty
\bibitem [{\citenamefont {Ashkin}(1970)}]{ashkin1970}%
  \BibitemOpen
  \bibfield  {author} {\bibinfo {author} {\bibfnamefont {A.}~\bibnamefont {Ashkin}},\ }\href {https://doi.org/10.1103/PhysRevLett.24.156} {\bibfield  {journal} {\bibinfo  {journal} {Phys. Rev. Lett.}\ }\textbf {\bibinfo {volume} {24}},\ \bibinfo {pages} {156} (\bibinfo {year} {1970})}\BibitemShut {NoStop}%
\bibitem [{\citenamefont {Ashkin}\ \emph {et~al.}(1986)\citenamefont {Ashkin}, \citenamefont {Dziedzic}, \citenamefont {Bjorkholm},\ and\ \citenamefont {Chu}}]{ashkin1986}%
  \BibitemOpen
  \bibfield  {author} {\bibinfo {author} {\bibfnamefont {A.}~\bibnamefont {Ashkin}}, \bibinfo {author} {\bibfnamefont {J.~M.}\ \bibnamefont {Dziedzic}}, \bibinfo {author} {\bibfnamefont {J.~E.}\ \bibnamefont {Bjorkholm}},\ and\ \bibinfo {author} {\bibfnamefont {S.}~\bibnamefont {Chu}},\ }\href {https://doi.org/10.1364/OL.11.000288} {\bibfield  {journal} {\bibinfo  {journal} {Opt. Lett.}\ }\textbf {\bibinfo {volume} {11}},\ \bibinfo {pages} {288} (\bibinfo {year} {1986})}\BibitemShut {NoStop}%
\bibitem [{\citenamefont {Ashkin}\ and\ \citenamefont {Dziedzic}(1987)}]{Ashkin1997}%
  \BibitemOpen
  \bibfield  {author} {\bibinfo {author} {\bibfnamefont {A.}~\bibnamefont {Ashkin}}\ and\ \bibinfo {author} {\bibfnamefont {J.~M.}\ \bibnamefont {Dziedzic}},\ }\href {https://doi.org/10.1126/science.3547653} {\bibfield  {journal} {\bibinfo  {journal} {Science}\ }\textbf {\bibinfo {volume} {235}},\ \bibinfo {pages} {1517} (\bibinfo {year} {1987})}\BibitemShut {NoStop}%
\bibitem [{\citenamefont {Guck}\ \emph {et~al.}(2001)\citenamefont {Guck}, \citenamefont {Ananthakrishnan}, \citenamefont {Mahmood}, \citenamefont {Moon}, \citenamefont {Cunningham},\ and\ \citenamefont {Kas}}]{Guck2001}%
  \BibitemOpen
  \bibfield  {author} {\bibinfo {author} {\bibfnamefont {J.}~\bibnamefont {Guck}}, \bibinfo {author} {\bibfnamefont {R.}~\bibnamefont {Ananthakrishnan}}, \bibinfo {author} {\bibfnamefont {H.}~\bibnamefont {Mahmood}}, \bibinfo {author} {\bibfnamefont {T.~J.}\ \bibnamefont {Moon}}, \bibinfo {author} {\bibfnamefont {C.~C.}\ \bibnamefont {Cunningham}},\ and\ \bibinfo {author} {\bibfnamefont {J.}~\bibnamefont {Kas}},\ }\href {https://doi.org/10.1016/S0006-3495(01)75740-2} {\bibfield  {journal} {\bibinfo  {journal} {Biophys. J.}\ }\textbf {\bibinfo {volume} {81}},\ \bibinfo {pages} {767} (\bibinfo {year} {2001})}\BibitemShut {NoStop}%
\bibitem [{\citenamefont {Grier}(2003)}]{Grier2003}%
  \BibitemOpen
  \bibfield  {author} {\bibinfo {author} {\bibfnamefont {D.~G.}\ \bibnamefont {Grier}},\ }\href {https://doi.org/10.1038/nature01935} {\bibfield  {journal} {\bibinfo  {journal} {Nature}\ }\textbf {\bibinfo {volume} {424}},\ \bibinfo {pages} {810} (\bibinfo {year} {2003})}\BibitemShut {NoStop}%
\bibitem [{\citenamefont {Neuman}\ and\ \citenamefont {Block}(2004)}]{Neuman2004}%
  \BibitemOpen
  \bibfield  {author} {\bibinfo {author} {\bibfnamefont {K.~C.}\ \bibnamefont {Neuman}}\ and\ \bibinfo {author} {\bibfnamefont {S.~M.}\ \bibnamefont {Block}},\ }\href {https://doi.org/10.1063/1.1785844} {\bibfield  {journal} {\bibinfo  {journal} {Rev. Sci. Instrum.}\ }\textbf {\bibinfo {volume} {75}},\ \bibinfo {pages} {2787} (\bibinfo {year} {2004})}\BibitemShut {NoStop}%
\bibitem [{\citenamefont {Juan}\ \emph {et~al.}(2011)\citenamefont {Juan}, \citenamefont {Righini},\ and\ \citenamefont {Quidant}}]{Juan2011}%
  \BibitemOpen
  \bibfield  {author} {\bibinfo {author} {\bibfnamefont {M.~L.}\ \bibnamefont {Juan}}, \bibinfo {author} {\bibfnamefont {M.}~\bibnamefont {Righini}},\ and\ \bibinfo {author} {\bibfnamefont {R.}~\bibnamefont {Quidant}},\ }\href {https://doi.org/10.1038/nphoton.2011.56} {\bibfield  {journal} {\bibinfo  {journal} {Nature Photonics}\ }\textbf {\bibinfo {volume} {5}},\ \bibinfo {pages} {349} (\bibinfo {year} {2011})}\BibitemShut {NoStop}%
\bibitem [{\citenamefont {Marago}\ \emph {et~al.}(2013)\citenamefont {Marago}, \citenamefont {Jones}, \citenamefont {Gucciardi}, \citenamefont {Volpe},\ and\ \citenamefont {Ferrari}}]{Marago2013}%
  \BibitemOpen
  \bibfield  {author} {\bibinfo {author} {\bibfnamefont {O.~M.}\ \bibnamefont {Marago}}, \bibinfo {author} {\bibfnamefont {P.~H.}\ \bibnamefont {Jones}}, \bibinfo {author} {\bibfnamefont {P.~G.}\ \bibnamefont {Gucciardi}}, \bibinfo {author} {\bibfnamefont {G.}~\bibnamefont {Volpe}},\ and\ \bibinfo {author} {\bibfnamefont {A.~C.}\ \bibnamefont {Ferrari}},\ }\href {https://doi.org/10.1038/NNANO.2013.208} {\bibfield  {journal} {\bibinfo  {journal} {Nat. Nanotechnol.}\ }\textbf {\bibinfo {volume} {8}},\ \bibinfo {pages} {807} (\bibinfo {year} {2013})}\BibitemShut {NoStop}%
\bibitem [{\citenamefont {Siviloglou}\ \emph {et~al.}(2007)\citenamefont {Siviloglou}, \citenamefont {Broky}, \citenamefont {Dogariu},\ and\ \citenamefont {Christodoulides}}]{airybeam2007}%
  \BibitemOpen
  \bibfield  {author} {\bibinfo {author} {\bibfnamefont {G.~A.}\ \bibnamefont {Siviloglou}}, \bibinfo {author} {\bibfnamefont {J.}~\bibnamefont {Broky}}, \bibinfo {author} {\bibfnamefont {A.}~\bibnamefont {Dogariu}},\ and\ \bibinfo {author} {\bibfnamefont {D.~N.}\ \bibnamefont {Christodoulides}},\ }\href {https://doi.org/10.1103/PhysRevLett.99.213901} {\bibfield  {journal} {\bibinfo  {journal} {Phys. Rev. Lett.}\ }\textbf {\bibinfo {volume} {99}},\ \bibinfo {pages} {213901} (\bibinfo {year} {2007})}\BibitemShut {NoStop}%
\bibitem [{\citenamefont {Baumgartl}\ \emph {et~al.}(2008)\citenamefont {Baumgartl}, \citenamefont {Mazilu},\ and\ \citenamefont {Dholakia}}]{airybeam2008}%
  \BibitemOpen
  \bibfield  {author} {\bibinfo {author} {\bibfnamefont {J.}~\bibnamefont {Baumgartl}}, \bibinfo {author} {\bibfnamefont {M.}~\bibnamefont {Mazilu}},\ and\ \bibinfo {author} {\bibfnamefont {K.}~\bibnamefont {Dholakia}},\ }\href {https://doi.org/10.1038/nphoton.2008.201} {\bibfield  {journal} {\bibinfo  {journal} {Nat. Photonics}\ }\textbf {\bibinfo {volume} {2}},\ \bibinfo {pages} {675} (\bibinfo {year} {2008})}\BibitemShut {NoStop}%
\bibitem [{\citenamefont {Durnin}\ \emph {et~al.}(1987)\citenamefont {Durnin}, \citenamefont {Miceli},\ and\ \citenamefont {Eberly}}]{Durnin1987}%
  \BibitemOpen
  \bibfield  {author} {\bibinfo {author} {\bibfnamefont {J.}~\bibnamefont {Durnin}}, \bibinfo {author} {\bibfnamefont {J.~J.}\ \bibnamefont {Miceli}},\ and\ \bibinfo {author} {\bibfnamefont {J.~H.}\ \bibnamefont {Eberly}},\ }\href {https://doi.org/10.1103/PhysRevLett.58.1499} {\bibfield  {journal} {\bibinfo  {journal} {Phys. Rev. Lett.}\ }\textbf {\bibinfo {volume} {58}},\ \bibinfo {pages} {1499} (\bibinfo {year} {1987})}\BibitemShut {NoStop}%
\bibitem [{\citenamefont {Chen}\ \emph {et~al.}(2011)\citenamefont {Chen}, \citenamefont {Ng}, \citenamefont {Lin},\ and\ \citenamefont {Chan}}]{pullingchen2011}%
  \BibitemOpen
  \bibfield  {author} {\bibinfo {author} {\bibfnamefont {J.}~\bibnamefont {Chen}}, \bibinfo {author} {\bibfnamefont {J.}~\bibnamefont {Ng}}, \bibinfo {author} {\bibfnamefont {Z.}~\bibnamefont {Lin}},\ and\ \bibinfo {author} {\bibfnamefont {C.~T.}\ \bibnamefont {Chan}},\ }\href {https://doi.org/10.1038/NPHOTON.2011.153} {\bibfield  {journal} {\bibinfo  {journal} {Nat. Photonics}\ }\textbf {\bibinfo {volume} {5}},\ \bibinfo {pages} {531} (\bibinfo {year} {2011})}\BibitemShut {NoStop}%
\bibitem [{\citenamefont {Novitsky}\ \emph {et~al.}(2011)\citenamefont {Novitsky}, \citenamefont {Qiu},\ and\ \citenamefont {Wang}}]{WOS:000297132200002}%
  \BibitemOpen
  \bibfield  {author} {\bibinfo {author} {\bibfnamefont {A.}~\bibnamefont {Novitsky}}, \bibinfo {author} {\bibfnamefont {C.-W.}\ \bibnamefont {Qiu}},\ and\ \bibinfo {author} {\bibfnamefont {H.}~\bibnamefont {Wang}},\ }\href {https://doi.org/10.1103/PhysRevLett.107.203601} {\bibfield  {journal} {\bibinfo  {journal} {Phys. Rev. Lett.}\ }\textbf {\bibinfo {volume} {107}},\ \bibinfo {pages} {203601} (\bibinfo {year} {2011})}\BibitemShut {NoStop}%
\bibitem [{\citenamefont {Curtis}\ \emph {et~al.}(2002)\citenamefont {Curtis}, \citenamefont {Koss},\ and\ \citenamefont {Grier}}]{holographictweezers2002}%
  \BibitemOpen
  \bibfield  {author} {\bibinfo {author} {\bibfnamefont {J.}~\bibnamefont {Curtis}}, \bibinfo {author} {\bibfnamefont {B.}~\bibnamefont {Koss}},\ and\ \bibinfo {author} {\bibfnamefont {D.}~\bibnamefont {Grier}},\ }\href {https://doi.org/10.1016/S0030-4018(02)01524-9} {\bibfield  {journal} {\bibinfo  {journal} {Opt. Commun.}\ }\textbf {\bibinfo {volume} {207}},\ \bibinfo {pages} {169} (\bibinfo {year} {2002})}\BibitemShut {NoStop}%
\bibitem [{\citenamefont {Yang}\ \emph {et~al.}(2021)\citenamefont {Yang}, \citenamefont {Ren}, \citenamefont {Chen}, \citenamefont {Arita},\ and\ \citenamefont {Rosales-Guzman}}]{Advphotonics2021}%
  \BibitemOpen
  \bibfield  {author} {\bibinfo {author} {\bibfnamefont {Y.}~\bibnamefont {Yang}}, \bibinfo {author} {\bibfnamefont {Y.-X.}\ \bibnamefont {Ren}}, \bibinfo {author} {\bibfnamefont {M.}~\bibnamefont {Chen}}, \bibinfo {author} {\bibfnamefont {Y.}~\bibnamefont {Arita}},\ and\ \bibinfo {author} {\bibfnamefont {C.}~\bibnamefont {Rosales-Guzman}},\ }\href {https://doi.org/10.1117/1.AP.3.3.034001} {\bibfield  {journal} {\bibinfo  {journal} {Adv. Photonics}\ }\textbf {\bibinfo {volume} {3}},\ \bibinfo {pages} {034001} (\bibinfo {year} {2021})}\BibitemShut {NoStop}%
\bibitem [{\citenamefont {Ng}\ \emph {et~al.}(2010)\citenamefont {Ng}, \citenamefont {Lin},\ and\ \citenamefont {Chan}}]{Vortex2010}%
  \BibitemOpen
  \bibfield  {author} {\bibinfo {author} {\bibfnamefont {J.}~\bibnamefont {Ng}}, \bibinfo {author} {\bibfnamefont {Z.}~\bibnamefont {Lin}},\ and\ \bibinfo {author} {\bibfnamefont {C.~T.}\ \bibnamefont {Chan}},\ }\href {https://doi.org/10.1103/PhysRevLett.104.103601} {\bibfield  {journal} {\bibinfo  {journal} {Phys. Rev. Lett.}\ }\textbf {\bibinfo {volume} {104}},\ \bibinfo {pages} {103601} (\bibinfo {year} {2010})}\BibitemShut {NoStop}%
\bibitem [{\citenamefont {Zhou}\ \emph {et~al.}(2022)\citenamefont {Zhou}, \citenamefont {Xu}, \citenamefont {Zhang} \emph {et~al.}}]{ImPoynting2022}%
  \BibitemOpen
  \bibfield  {author} {\bibinfo {author} {\bibfnamefont {Y.}~\bibnamefont {Zhou}}, \bibinfo {author} {\bibfnamefont {X.}~\bibnamefont {Xu}}, \bibinfo {author} {\bibfnamefont {Y.}~\bibnamefont {Zhang}}, \emph {et~al.},\ }\href {https://doi.org/10.1073/pnas.2209721119} {\bibfield  {journal} {\bibinfo  {journal} {Proc. Natl. Acad. Sci. U. S. A.}\ }\textbf {\bibinfo {volume} {119}},\ \bibinfo {pages} {e2209721119} (\bibinfo {year} {2022})}\BibitemShut {NoStop}%
\bibitem [{\citenamefont {Phillips}\ \emph {et~al.}(2014)\citenamefont {Phillips}, \citenamefont {Padgett}, \citenamefont {Hanna}, \citenamefont {Ho}, \citenamefont {Carberry}, \citenamefont {Miles},\ and\ \citenamefont {Simpson}}]{WOS:000335011700015}%
  \BibitemOpen
  \bibfield  {author} {\bibinfo {author} {\bibfnamefont {D.~B.}\ \bibnamefont {Phillips}}, \bibinfo {author} {\bibfnamefont {M.~J.}\ \bibnamefont {Padgett}}, \bibinfo {author} {\bibfnamefont {S.}~\bibnamefont {Hanna}}, \bibinfo {author} {\bibfnamefont {Y.~L.~D.}\ \bibnamefont {Ho}}, \bibinfo {author} {\bibfnamefont {D.~M.}\ \bibnamefont {Carberry}}, \bibinfo {author} {\bibfnamefont {M.~J.}\ \bibnamefont {Miles}},\ and\ \bibinfo {author} {\bibfnamefont {S.~H.}\ \bibnamefont {Simpson}},\ }\href {https://doi.org/10.1038/nphoton.2014.74} {\bibfield  {journal} {\bibinfo  {journal} {Nat. Photonics}\ }\textbf {\bibinfo {volume} {8}},\ \bibinfo {pages} {400} (\bibinfo {year} {2014})}\BibitemShut {NoStop}%
\bibitem [{\citenamefont {Li}\ \emph {et~al.}(2019)\citenamefont {Li}, \citenamefont {Chen}, \citenamefont {Lin},\ and\ \citenamefont {Ng}}]{SciAdvpulling2019}%
  \BibitemOpen
  \bibfield  {author} {\bibinfo {author} {\bibfnamefont {X.}~\bibnamefont {Li}}, \bibinfo {author} {\bibfnamefont {J.}~\bibnamefont {Chen}}, \bibinfo {author} {\bibfnamefont {Z.}~\bibnamefont {Lin}},\ and\ \bibinfo {author} {\bibfnamefont {J.}~\bibnamefont {Ng}},\ }\href {https://doi.org/10.1126/sciadv.aau7814} {\bibfield  {journal} {\bibinfo  {journal} {Sci. Adv.}\ }\textbf {\bibinfo {volume} {5}},\ \bibinfo {pages} {eaau7814} (\bibinfo {year} {2019})}\BibitemShut {NoStop}%
\bibitem [{\citenamefont {Swartzlander}\ \emph {et~al.}(2011)\citenamefont {Swartzlander}, \citenamefont {Peterson}, \citenamefont {Artusio-Glimpse},\ and\ \citenamefont {Raisanen}}]{opticallift2011}%
  \BibitemOpen
  \bibfield  {author} {\bibinfo {author} {\bibfnamefont {G.~A.}\ \bibnamefont {Swartzlander}, \bibfnamefont {Jr.}}, \bibinfo {author} {\bibfnamefont {T.~J.}\ \bibnamefont {Peterson}}, \bibinfo {author} {\bibfnamefont {A.~B.}\ \bibnamefont {Artusio-Glimpse}},\ and\ \bibinfo {author} {\bibfnamefont {A.~D.}\ \bibnamefont {Raisanen}},\ }\href {https://doi.org/10.1038/NPHOTON.2010.266} {\bibfield  {journal} {\bibinfo  {journal} {Nat. Photonics}\ }\textbf {\bibinfo {volume} {5}},\ \bibinfo {pages} {48} (\bibinfo {year} {2011})}\BibitemShut {NoStop}%
\bibitem [{\citenamefont {Mao}\ \emph {et~al.}(2025)\citenamefont {Mao}, \citenamefont {Toftul}, \citenamefont {Balendhran}, \citenamefont {Taha}, \citenamefont {Kivshar},\ and\ \citenamefont {Kruk}}]{PhaseChang2025}%
  \BibitemOpen
  \bibfield  {author} {\bibinfo {author} {\bibfnamefont {L.}~\bibnamefont {Mao}}, \bibinfo {author} {\bibfnamefont {I.}~\bibnamefont {Toftul}}, \bibinfo {author} {\bibfnamefont {S.}~\bibnamefont {Balendhran}}, \bibinfo {author} {\bibfnamefont {M.}~\bibnamefont {Taha}}, \bibinfo {author} {\bibfnamefont {Y.}~\bibnamefont {Kivshar}},\ and\ \bibinfo {author} {\bibfnamefont {S.}~\bibnamefont {Kruk}},\ }\href {https://doi.org/10.1002/lpor.202400767} {\bibfield  {journal} {\bibinfo  {journal} {Laser Photon. Rev.}\ }\textbf {\bibinfo {volume} {19}},\ \bibinfo {pages} {2400767} (\bibinfo {year} {2025})}\BibitemShut {NoStop}%
\bibitem [{\citenamefont {Gao}\ \emph {et~al.}(2020)\citenamefont {Gao}, \citenamefont {Wang}, \citenamefont {He}, \citenamefont {Xu}, \citenamefont {Zhu}, \citenamefont {Hu}, \citenamefont {Hu}, \citenamefont {Li},\ and\ \citenamefont {Hu}}]{Janusparitcle}%
  \BibitemOpen
  \bibfield  {author} {\bibinfo {author} {\bibfnamefont {X.}~\bibnamefont {Gao}}, \bibinfo {author} {\bibfnamefont {Y.}~\bibnamefont {Wang}}, \bibinfo {author} {\bibfnamefont {X.}~\bibnamefont {He}}, \bibinfo {author} {\bibfnamefont {M.}~\bibnamefont {Xu}}, \bibinfo {author} {\bibfnamefont {J.}~\bibnamefont {Zhu}}, \bibinfo {author} {\bibfnamefont {X.}~\bibnamefont {Hu}}, \bibinfo {author} {\bibfnamefont {X.}~\bibnamefont {Hu}}, \bibinfo {author} {\bibfnamefont {H.}~\bibnamefont {Li}},\ and\ \bibinfo {author} {\bibfnamefont {C.}~\bibnamefont {Hu}},\ }\href {https://doi.org/10.1002/smtd.202000565} {\bibfield  {journal} {\bibinfo  {journal} {Small Methods}\ }\textbf {\bibinfo {volume} {4}},\ \bibinfo {pages} {2000565} (\bibinfo {year} {2020})}\BibitemShut {NoStop}%
\bibitem [{\citenamefont {Karpinski}(2022)}]{WOS:000712284300001}%
  \BibitemOpen
  \bibfield  {author} {\bibinfo {author} {\bibfnamefont {P.}~\bibnamefont {Karpinski}},\ }\href {https://doi.org/doi:10.1002/adom.202101592} {\bibfield  {journal} {\bibinfo  {journal} {Adv. Opt. Mater.}\ }\textbf {\bibinfo {volume} {10}},\ \bibinfo {pages} {2101592} (\bibinfo {year} {2022})}\BibitemShut {NoStop}%
\bibitem [{\citenamefont {Rodrigo}\ \emph {et~al.}(2024)\citenamefont {Rodrigo}, \citenamefont {Alieva}, \citenamefont {Manzaneda-Gonzalez},\ and\ \citenamefont {Guerrero-Martinez}}]{WOS:001323979500001}%
  \BibitemOpen
  \bibfield  {author} {\bibinfo {author} {\bibfnamefont {J.~A.}\ \bibnamefont {Rodrigo}}, \bibinfo {author} {\bibfnamefont {T.}~\bibnamefont {Alieva}}, \bibinfo {author} {\bibfnamefont {V.}~\bibnamefont {Manzaneda-Gonzalez}},\ and\ \bibinfo {author} {\bibfnamefont {A.}~\bibnamefont {Guerrero-Martinez}},\ }\href {https://doi.org/10.1021/acsnano.4c10264} {\bibfield  {journal} {\bibinfo  {journal} {ACS Nano}\ }\textbf {\bibinfo {volume} {18}},\ \bibinfo {pages} {27738} (\bibinfo {year} {2024})}\BibitemShut {NoStop}%
\bibitem [{\citenamefont {Arai}\ \emph {et~al.}(1999)\citenamefont {Arai}, \citenamefont {Yasuda}, \citenamefont {Akashi}, \citenamefont {Harada}, \citenamefont {Miyata}, \citenamefont {Kinosita},\ and\ \citenamefont {Itoh}}]{molecularknot1999}%
  \BibitemOpen
  \bibfield  {author} {\bibinfo {author} {\bibfnamefont {Y.}~\bibnamefont {Arai}}, \bibinfo {author} {\bibfnamefont {R.}~\bibnamefont {Yasuda}}, \bibinfo {author} {\bibfnamefont {K.}~\bibnamefont {Akashi}}, \bibinfo {author} {\bibfnamefont {Y.}~\bibnamefont {Harada}}, \bibinfo {author} {\bibfnamefont {H.}~\bibnamefont {Miyata}}, \bibinfo {author} {\bibfnamefont {K.}~\bibnamefont {Kinosita}},\ and\ \bibinfo {author} {\bibfnamefont {H.}~\bibnamefont {Itoh}},\ }\href {https://doi.org/10.1038/20894} {\bibfield  {journal} {\bibinfo  {journal} {Nature}\ }\textbf {\bibinfo {volume} {399}},\ \bibinfo {pages} {446} (\bibinfo {year} {1999})}\BibitemShut {NoStop}%
\bibitem [{\citenamefont {Bao}\ \emph {et~al.}(2003)\citenamefont {Bao}, \citenamefont {Lee},\ and\ \citenamefont {Quake}}]{WOS:000187719300038}%
  \BibitemOpen
  \bibfield  {author} {\bibinfo {author} {\bibfnamefont {X.}~\bibnamefont {Bao}}, \bibinfo {author} {\bibfnamefont {H.}~\bibnamefont {Lee}},\ and\ \bibinfo {author} {\bibfnamefont {S.}~\bibnamefont {Quake}},\ }\href {https://doi.org/10.1103/PhysRevLett.91.265506} {\bibfield  {journal} {\bibinfo  {journal} {Phys. Rev. Lett.}\ }\textbf {\bibinfo {volume} {91}},\ \bibinfo {pages} {265506} (\bibinfo {year} {2003})}\BibitemShut {NoStop}%
\bibitem [{\citenamefont {Millen}\ \emph {et~al.}(2020)\citenamefont {Millen}, \citenamefont {Monteiro}, \citenamefont {Pettit},\ and\ \citenamefont {Vamivakas}}]{WOS:000508458700001}%
  \BibitemOpen
  \bibfield  {author} {\bibinfo {author} {\bibfnamefont {J.}~\bibnamefont {Millen}}, \bibinfo {author} {\bibfnamefont {T.~S.}\ \bibnamefont {Monteiro}}, \bibinfo {author} {\bibfnamefont {R.}~\bibnamefont {Pettit}},\ and\ \bibinfo {author} {\bibfnamefont {A.~N.}\ \bibnamefont {Vamivakas}},\ }\href {https://doi.org/10.1088/1361-6633/ab6100} {\bibfield  {journal} {\bibinfo  {journal} {Rep. Prog. Phys.}\ }\textbf {\bibinfo {volume} {83}},\ \bibinfo {pages} {026401} (\bibinfo {year} {2020})}\BibitemShut {NoStop}%
\bibitem [{\citenamefont {Xin}\ \emph {et~al.}(2020)\citenamefont {Xin}, \citenamefont {Li}, \citenamefont {Liu}, \citenamefont {Zhang}, \citenamefont {Xiao},\ and\ \citenamefont {Li}}]{WOS:000552283000001}%
  \BibitemOpen
  \bibfield  {author} {\bibinfo {author} {\bibfnamefont {H.}~\bibnamefont {Xin}}, \bibinfo {author} {\bibfnamefont {Y.}~\bibnamefont {Li}}, \bibinfo {author} {\bibfnamefont {Y.-C.}\ \bibnamefont {Liu}}, \bibinfo {author} {\bibfnamefont {Y.}~\bibnamefont {Zhang}}, \bibinfo {author} {\bibfnamefont {Y.-F.}\ \bibnamefont {Xiao}},\ and\ \bibinfo {author} {\bibfnamefont {B.}~\bibnamefont {Li}},\ }\href {https://doi.org/10.1002/adma.202001994} {\bibfield  {journal} {\bibinfo  {journal} {Adv. Mater.}\ }\textbf {\bibinfo {volume} {32}},\ \bibinfo {pages} {2001994} (\bibinfo {year} {2020})}\BibitemShut {NoStop}%
\bibitem [{\citenamefont {Li}\ \emph {et~al.}(2024)\citenamefont {Li}, \citenamefont {Cao}, \citenamefont {Feng}, \citenamefont {Shi}, \citenamefont {Shi}, \citenamefont {Chen}, \citenamefont {Gao}, \citenamefont {Zhu}, \citenamefont {Tang}, \citenamefont {Sun}, \citenamefont {Qiu},\ and\ \citenamefont {Ding}}]{WOS:001269001300006}%
  \BibitemOpen
  \bibfield  {author} {\bibinfo {author} {\bibfnamefont {H.}~\bibnamefont {Li}}, \bibinfo {author} {\bibfnamefont {Y.}~\bibnamefont {Cao}}, \bibinfo {author} {\bibfnamefont {R.}~\bibnamefont {Feng}}, \bibinfo {author} {\bibfnamefont {B.}~\bibnamefont {Shi}}, \bibinfo {author} {\bibfnamefont {Y.}~\bibnamefont {Shi}}, \bibinfo {author} {\bibfnamefont {Y.}~\bibnamefont {Chen}}, \bibinfo {author} {\bibfnamefont {D.}~\bibnamefont {Gao}}, \bibinfo {author} {\bibfnamefont {T.}~\bibnamefont {Zhu}}, \bibinfo {author} {\bibfnamefont {D.}~\bibnamefont {Tang}}, \bibinfo {author} {\bibfnamefont {F.}~\bibnamefont {Sun}}, \bibinfo {author} {\bibfnamefont {C.-W.}\ \bibnamefont {Qiu}},\ and\ \bibinfo {author} {\bibfnamefont {W.}~\bibnamefont {Ding}},\ }\href {https://doi.org/10.1103/PhysRevLett.132.253802} {\bibfield  {journal} {\bibinfo  {journal} {Phys. Rev. Lett.}\ }\textbf {\bibinfo {volume} {132}},\ \bibinfo {pages} {253802} (\bibinfo {year} {2024})}\BibitemShut {NoStop}%
\bibitem [{\citenamefont {Jokisch}\ \emph {et~al.}(2024)\citenamefont {Jokisch}, \citenamefont {Gotzsche}, \citenamefont {Kristensen}, \citenamefont {Wubs}, \citenamefont {Sigmund},\ and\ \citenamefont {Christiansen}}]{WOS:001319882200001}%
  \BibitemOpen
  \bibfield  {author} {\bibinfo {author} {\bibfnamefont {B.~M. d.~A.}\ \bibnamefont {Jokisch}}, \bibinfo {author} {\bibfnamefont {B.~F.}\ \bibnamefont {Gotzsche}}, \bibinfo {author} {\bibfnamefont {P.~T.}\ \bibnamefont {Kristensen}}, \bibinfo {author} {\bibfnamefont {M.}~\bibnamefont {Wubs}}, \bibinfo {author} {\bibfnamefont {O.}~\bibnamefont {Sigmund}},\ and\ \bibinfo {author} {\bibfnamefont {R.~E.}\ \bibnamefont {Christiansen}},\ }\href {https://doi.org/10.1021/acsphotonics.4c01060} {\bibfield  {journal} {\bibinfo  {journal} {ACS Photonics}\ }\textbf {\bibinfo {volume} {11}},\ \bibinfo {pages} {5118} (\bibinfo {year} {2024})}\BibitemShut {NoStop}%
\bibitem [{\citenamefont {Koya}\ \emph {et~al.}(2020)\citenamefont {Koya}, \citenamefont {Cunha}, \citenamefont {Guo}, \citenamefont {Toma}, \citenamefont {Garoli}, \citenamefont {Wang}, \citenamefont {Juodkazis}, \citenamefont {Cojoc},\ and\ \citenamefont {Zaccaria}}]{WOS:000516663000001}%
  \BibitemOpen
  \bibfield  {author} {\bibinfo {author} {\bibfnamefont {A.~N.}\ \bibnamefont {Koya}}, \bibinfo {author} {\bibfnamefont {J.}~\bibnamefont {Cunha}}, \bibinfo {author} {\bibfnamefont {T.-L.}\ \bibnamefont {Guo}}, \bibinfo {author} {\bibfnamefont {A.}~\bibnamefont {Toma}}, \bibinfo {author} {\bibfnamefont {D.}~\bibnamefont {Garoli}}, \bibinfo {author} {\bibfnamefont {T.}~\bibnamefont {Wang}}, \bibinfo {author} {\bibfnamefont {S.}~\bibnamefont {Juodkazis}}, \bibinfo {author} {\bibfnamefont {D.}~\bibnamefont {Cojoc}},\ and\ \bibinfo {author} {\bibfnamefont {R.~P.}\ \bibnamefont {Zaccaria}},\ }\href {https://doi.org/10.1002/adom.201901481} {\bibfield  {journal} {\bibinfo  {journal} {Adv. Opt. Mater.}\ }\textbf {\bibinfo {volume} {8}},\ \bibinfo {pages} {1901481} (\bibinfo {year} {2020})}\BibitemShut {NoStop}%
\bibitem [{\citenamefont {Ren}\ \emph {et~al.}(2021)\citenamefont {Ren}, \citenamefont {Chen}, \citenamefont {He}, \citenamefont {Zhang}, \citenamefont {Qi},\ and\ \citenamefont {Yan}}]{WOS:000645436800017}%
  \BibitemOpen
  \bibfield  {author} {\bibinfo {author} {\bibfnamefont {Y.}~\bibnamefont {Ren}}, \bibinfo {author} {\bibfnamefont {Q.}~\bibnamefont {Chen}}, \bibinfo {author} {\bibfnamefont {M.}~\bibnamefont {He}}, \bibinfo {author} {\bibfnamefont {X.}~\bibnamefont {Zhang}}, \bibinfo {author} {\bibfnamefont {H.}~\bibnamefont {Qi}},\ and\ \bibinfo {author} {\bibfnamefont {Y.}~\bibnamefont {Yan}},\ }\href {https://doi.org/10.1021/acsnano.1c00466} {\bibfield  {journal} {\bibinfo  {journal} {ACS Nano}\ }\textbf {\bibinfo {volume} {15}},\ \bibinfo {pages} {6105} (\bibinfo {year} {2021})}\BibitemShut {NoStop}%
\bibitem [{\citenamefont {Zhang}\ \emph {et~al.}(2021)\citenamefont {Zhang}, \citenamefont {Min}, \citenamefont {Dou}, \citenamefont {Wang}, \citenamefont {Urbach}, \citenamefont {Somekh},\ and\ \citenamefont {Yuan}}]{WOS:000630005400001}%
  \BibitemOpen
  \bibfield  {author} {\bibinfo {author} {\bibfnamefont {Y.}~\bibnamefont {Zhang}}, \bibinfo {author} {\bibfnamefont {C.}~\bibnamefont {Min}}, \bibinfo {author} {\bibfnamefont {X.}~\bibnamefont {Dou}}, \bibinfo {author} {\bibfnamefont {X.}~\bibnamefont {Wang}}, \bibinfo {author} {\bibfnamefont {H.~P.}\ \bibnamefont {Urbach}}, \bibinfo {author} {\bibfnamefont {M.~G.}\ \bibnamefont {Somekh}},\ and\ \bibinfo {author} {\bibfnamefont {X.}~\bibnamefont {Yuan}},\ }\href {https://doi.org/10.1038/s41377-021-00474-0} {\bibfield  {journal} {\bibinfo  {journal} {Light-Sci. Appl.}\ }\textbf {\bibinfo {volume} {10}},\ \bibinfo {pages} {59} (\bibinfo {year} {2021})}\BibitemShut {NoStop}%
\bibitem [{\citenamefont {Wang}\ \emph {et~al.}(2025)\citenamefont {Wang}, \citenamefont {Li}, \citenamefont {Xia} \emph {et~al.}}]{WOS:001456600400001}%
  \BibitemOpen
  \bibfield  {author} {\bibinfo {author} {\bibfnamefont {R.}~\bibnamefont {Wang}}, \bibinfo {author} {\bibfnamefont {W.}~\bibnamefont {Li}}, \bibinfo {author} {\bibfnamefont {Z.}~\bibnamefont {Xia}}, \emph {et~al.},\ }\href {https://doi.org/10.1038/s41377-025-01801-5} {\bibfield  {journal} {\bibinfo  {journal} {Light-Sci. Appl.}\ }\textbf {\bibinfo {volume} {14}},\ \bibinfo {pages} {146} (\bibinfo {year} {2025})}\BibitemShut {NoStop}%
\bibitem [{\citenamefont {Koya}\ \emph {et~al.}(2023)\citenamefont {Koya}, \citenamefont {Li},\ and\ \citenamefont {Li}}]{WOS:001110458000005}%
  \BibitemOpen
  \bibfield  {author} {\bibinfo {author} {\bibfnamefont {A.~N.}\ \bibnamefont {Koya}}, \bibinfo {author} {\bibfnamefont {L.}~\bibnamefont {Li}},\ and\ \bibinfo {author} {\bibfnamefont {W.}~\bibnamefont {Li}},\ }\href {https://doi.org/10.1063/5.0178300} {\bibfield  {journal} {\bibinfo  {journal} {Appl. Phys. Lett.}\ }\textbf {\bibinfo {volume} {123}},\ \bibinfo {pages} {221107} (\bibinfo {year} {2023})}\BibitemShut {NoStop}%
\bibitem [{\citenamefont {Kajorndejnukul}\ \emph {et~al.}(2013)\citenamefont {Kajorndejnukul}, \citenamefont {Ding}, \citenamefont {Sukhov}, \citenamefont {Qiu},\ and\ \citenamefont {Dogariu}}]{NatPhotLinearmom2013}%
  \BibitemOpen
  \bibfield  {author} {\bibinfo {author} {\bibfnamefont {V.}~\bibnamefont {Kajorndejnukul}}, \bibinfo {author} {\bibfnamefont {W.}~\bibnamefont {Ding}}, \bibinfo {author} {\bibfnamefont {S.}~\bibnamefont {Sukhov}}, \bibinfo {author} {\bibfnamefont {C.-W.}\ \bibnamefont {Qiu}},\ and\ \bibinfo {author} {\bibfnamefont {A.}~\bibnamefont {Dogariu}},\ }\href {https://doi.org/10.1038/NPHOTON.2013.192} {\bibfield  {journal} {\bibinfo  {journal} {Nat. Photonics}\ }\textbf {\bibinfo {volume} {7}},\ \bibinfo {pages} {787} (\bibinfo {year} {2013})}\BibitemShut {NoStop}%
\bibitem [{\citenamefont {Li}\ \emph {et~al.}(2025)\citenamefont {Li}, \citenamefont {Zhu}, \citenamefont {Cao} \emph {et~al.}}]{WOS:001297264800001}%
  \BibitemOpen
  \bibfield  {author} {\bibinfo {author} {\bibfnamefont {H.}~\bibnamefont {Li}}, \bibinfo {author} {\bibfnamefont {T.}~\bibnamefont {Zhu}}, \bibinfo {author} {\bibfnamefont {Y.}~\bibnamefont {Cao}}, \emph {et~al.},\ }\href {https://doi.org/10.1002/lpor.202400330} {\bibfield  {journal} {\bibinfo  {journal} {Laser Photon. Rev.}\ }\textbf {\bibinfo {volume} {19}},\ \bibinfo {pages} {202400330} (\bibinfo {year} {2025})}\BibitemShut {NoStop}%
\bibitem [{\citenamefont {Lee}\ \emph {et~al.}(2020)\citenamefont {Lee}, \citenamefont {Huang},\ and\ \citenamefont {Luo}}]{WOS:000536569900002}%
  \BibitemOpen
  \bibfield  {author} {\bibinfo {author} {\bibfnamefont {E.}~\bibnamefont {Lee}}, \bibinfo {author} {\bibfnamefont {D.}~\bibnamefont {Huang}},\ and\ \bibinfo {author} {\bibfnamefont {T.}~\bibnamefont {Luo}},\ }\href {https://doi.org/10.1038/s41467-020-16267-9} {\bibfield  {journal} {\bibinfo  {journal} {Nat. Commun.}\ }\textbf {\bibinfo {volume} {11}},\ \bibinfo {pages} {2404} (\bibinfo {year} {2020})}\BibitemShut {NoStop}%
\bibitem [{\citenamefont {Lee}\ and\ \citenamefont {Luo}(2020)}]{WOS:000537235300017}%
  \BibitemOpen
  \bibfield  {author} {\bibinfo {author} {\bibfnamefont {E.}~\bibnamefont {Lee}}\ and\ \bibinfo {author} {\bibfnamefont {T.}~\bibnamefont {Luo}},\ }\href {https://doi.org/10.1126/sciadv.aaz3646} {\bibfield  {journal} {\bibinfo  {journal} {Sci. Adv.}\ }\textbf {\bibinfo {volume} {6}},\ \bibinfo {pages} {eaaz3646} (\bibinfo {year} {2020})}\BibitemShut {NoStop}%
\bibitem [{\citenamefont {Shi}\ \emph {et~al.}(2022)\citenamefont {Shi}, \citenamefont {Song}, \citenamefont {Toftul}, \citenamefont {Zhu}, \citenamefont {Yu}, \citenamefont {Zhu}, \citenamefont {Tsai}, \citenamefont {Kivshar},\ and\ \citenamefont {Liu}}]{WOS:000841997700001}%
  \BibitemOpen
  \bibfield  {author} {\bibinfo {author} {\bibfnamefont {Y.}~\bibnamefont {Shi}}, \bibinfo {author} {\bibfnamefont {Q.}~\bibnamefont {Song}}, \bibinfo {author} {\bibfnamefont {I.}~\bibnamefont {Toftul}}, \bibinfo {author} {\bibfnamefont {T.}~\bibnamefont {Zhu}}, \bibinfo {author} {\bibfnamefont {Y.}~\bibnamefont {Yu}}, \bibinfo {author} {\bibfnamefont {W.}~\bibnamefont {Zhu}}, \bibinfo {author} {\bibfnamefont {D.~P.}\ \bibnamefont {Tsai}}, \bibinfo {author} {\bibfnamefont {Y.}~\bibnamefont {Kivshar}},\ and\ \bibinfo {author} {\bibfnamefont {A.~Q.}\ \bibnamefont {Liu}},\ }\href {https://doi.org/10.1063/5.0091280} {\bibfield  {journal} {\bibinfo  {journal} {Appl. Phys. Rev.}\ }\textbf {\bibinfo {volume} {9}},\ \bibinfo {pages} {031303} (\bibinfo {year} {2022})}\BibitemShut {NoStop}%
\bibitem [{\citenamefont {Hsu}\ \emph {et~al.}(2022)\citenamefont {Hsu}, \citenamefont {Zhu}, \citenamefont {Thiele}, \citenamefont {Brown}, \citenamefont {Papp}, \citenamefont {Agrawal},\ and\ \citenamefont {Regal}}]{WOS:000841893000001}%
  \BibitemOpen
  \bibfield  {author} {\bibinfo {author} {\bibfnamefont {T.-W.}\ \bibnamefont {Hsu}}, \bibinfo {author} {\bibfnamefont {W.}~\bibnamefont {Zhu}}, \bibinfo {author} {\bibfnamefont {T.}~\bibnamefont {Thiele}}, \bibinfo {author} {\bibfnamefont {M.~O.}\ \bibnamefont {Brown}}, \bibinfo {author} {\bibfnamefont {S.~B.}\ \bibnamefont {Papp}}, \bibinfo {author} {\bibfnamefont {A.}~\bibnamefont {Agrawal}},\ and\ \bibinfo {author} {\bibfnamefont {C.~A.}\ \bibnamefont {Regal}},\ }\href {https://doi.org/10.1103/PRXQuantum.3.030316} {\bibfield  {journal} {\bibinfo  {journal} {PRX Quantum}\ }\textbf {\bibinfo {volume} {3}},\ \bibinfo {pages} {030316} (\bibinfo {year} {2022})}\BibitemShut {NoStop}%
\bibitem [{\citenamefont {Lin}\ \emph {et~al.}(2009)\citenamefont {Lin}, \citenamefont {Hu}, \citenamefont {Kimerling},\ and\ \citenamefont {Crozier}}]{WOS:000271374600068}%
  \BibitemOpen
  \bibfield  {author} {\bibinfo {author} {\bibfnamefont {S.}~\bibnamefont {Lin}}, \bibinfo {author} {\bibfnamefont {J.}~\bibnamefont {Hu}}, \bibinfo {author} {\bibfnamefont {L.}~\bibnamefont {Kimerling}},\ and\ \bibinfo {author} {\bibfnamefont {K.}~\bibnamefont {Crozier}},\ }\href {https://doi.org/10.1364/OL.34.003451} {\bibfield  {journal} {\bibinfo  {journal} {Opt. Lett.}\ }\textbf {\bibinfo {volume} {34}},\ \bibinfo {pages} {3451} (\bibinfo {year} {2009})}\BibitemShut {NoStop}%
\bibitem [{\citenamefont {Jin}\ \emph {et~al.}(2021)\citenamefont {Jin}, \citenamefont {Xu}, \citenamefont {Dong},\ and\ \citenamefont {Liu}}]{WOS:000731589200001}%
  \BibitemOpen
  \bibfield  {author} {\bibinfo {author} {\bibfnamefont {R.}~\bibnamefont {Jin}}, \bibinfo {author} {\bibfnamefont {Y.}~\bibnamefont {Xu}}, \bibinfo {author} {\bibfnamefont {Z.-G.}\ \bibnamefont {Dong}},\ and\ \bibinfo {author} {\bibfnamefont {Y.}~\bibnamefont {Liu}},\ }\href {https://doi.org/10.1021/acs.nanolett.1c03772} {\bibfield  {journal} {\bibinfo  {journal} {Nano Lett.}\ }\textbf {\bibinfo {volume} {21}},\ \bibinfo {pages} {10431} (\bibinfo {year} {2021})}\BibitemShut {NoStop}%
\bibitem [{\citenamefont {Zhang}\ \emph {et~al.}(2025)\citenamefont {Zhang}, \citenamefont {He}, \citenamefont {Li}, \citenamefont {Lu}, \citenamefont {Lai}, \citenamefont {Song}, \citenamefont {Wang}, \citenamefont {Shi}, \citenamefont {Wei},\ and\ \citenamefont {Cheng}}]{WOS:001456028800001}%
  \BibitemOpen
  \bibfield  {author} {\bibinfo {author} {\bibfnamefont {J.}~\bibnamefont {Zhang}}, \bibinfo {author} {\bibfnamefont {T.}~\bibnamefont {He}}, \bibinfo {author} {\bibfnamefont {C.}~\bibnamefont {Li}}, \bibinfo {author} {\bibfnamefont {C.}~\bibnamefont {Lu}}, \bibinfo {author} {\bibfnamefont {C.}~\bibnamefont {Lai}}, \bibinfo {author} {\bibfnamefont {Q.}~\bibnamefont {Song}}, \bibinfo {author} {\bibfnamefont {Z.}~\bibnamefont {Wang}}, \bibinfo {author} {\bibfnamefont {Y.}~\bibnamefont {Shi}}, \bibinfo {author} {\bibfnamefont {Z.}~\bibnamefont {Wei}},\ and\ \bibinfo {author} {\bibfnamefont {X.}~\bibnamefont {Cheng}},\ }\href {https://doi.org/10.1021/acs.nanolett.5c00406} {\bibfield  {journal} {\bibinfo  {journal} {Nano Lett.}\ }\textbf {\bibinfo {volume} {25}},\ \bibinfo {pages} {6539} (\bibinfo {year} {2025})}\BibitemShut {NoStop}%
\bibitem [{\citenamefont {Xiao}\ \emph {et~al.}(2023)\citenamefont {Xiao}, \citenamefont {Plaskocinski}, \citenamefont {Biabanifard}, \citenamefont {Persheyev},\ and\ \citenamefont {Di~Falco}}]{WOS:000954618400001}%
  \BibitemOpen
  \bibfield  {author} {\bibinfo {author} {\bibfnamefont {J.}~\bibnamefont {Xiao}}, \bibinfo {author} {\bibfnamefont {T.}~\bibnamefont {Plaskocinski}}, \bibinfo {author} {\bibfnamefont {M.}~\bibnamefont {Biabanifard}}, \bibinfo {author} {\bibfnamefont {S.}~\bibnamefont {Persheyev}},\ and\ \bibinfo {author} {\bibfnamefont {A.}~\bibnamefont {Di~Falco}},\ }\href {https://doi.org/10.1021/acsphotonics.2c01986} {\bibfield  {journal} {\bibinfo  {journal} {ACS Photonics}\ }\textbf {\bibinfo {volume} {10}},\ \bibinfo {pages} {1341} (\bibinfo {year} {2023})}\BibitemShut {NoStop}%
\bibitem [{\citenamefont {Li}\ \emph {et~al.}(2020)\citenamefont {Li}, \citenamefont {Cao}, \citenamefont {Shi} \emph {et~al.}}]{topology2020}%
  \BibitemOpen
  \bibfield  {author} {\bibinfo {author} {\bibfnamefont {H.}~\bibnamefont {Li}}, \bibinfo {author} {\bibfnamefont {Y.}~\bibnamefont {Cao}}, \bibinfo {author} {\bibfnamefont {B.}~\bibnamefont {Shi}}, \emph {et~al.},\ }\href {https://doi.org/10.1103/PhysRevLett.124.143901} {\bibfield  {journal} {\bibinfo  {journal} {Phys. Rev. Lett.}\ }\textbf {\bibinfo {volume} {124}},\ \bibinfo {pages} {143901} (\bibinfo {year} {2020})}\BibitemShut {NoStop}%
\bibitem [{\citenamefont {Rubinsztein-Dunlop}\ \emph {et~al.}(2017)\citenamefont {Rubinsztein-Dunlop}, \citenamefont {Forbes}, \citenamefont {Berry} \emph {et~al.}}]{structuredlightJOpt2017}%
  \BibitemOpen
  \bibfield  {author} {\bibinfo {author} {\bibfnamefont {H.}~\bibnamefont {Rubinsztein-Dunlop}}, \bibinfo {author} {\bibfnamefont {A.}~\bibnamefont {Forbes}}, \bibinfo {author} {\bibfnamefont {M.~V.}\ \bibnamefont {Berry}}, \emph {et~al.},\ }\href {https://doi.org/10.1088/2040-8978/19/1/013001} {\bibfield  {journal} {\bibinfo  {journal} {J. Opt.}\ }\textbf {\bibinfo {volume} {19}},\ \bibinfo {pages} {013001} (\bibinfo {year} {2017})}\BibitemShut {NoStop}%
\bibitem [{\citenamefont {Forbes}(2019)}]{structuredlight2019}%
  \BibitemOpen
  \bibfield  {author} {\bibinfo {author} {\bibfnamefont {A.}~\bibnamefont {Forbes}},\ }\href {https://doi.org/10.1002/lpor.201900140} {\bibfield  {journal} {\bibinfo  {journal} {Laser Photon. Rev.}\ }\textbf {\bibinfo {volume} {13}},\ \bibinfo {pages} {201900140} (\bibinfo {year} {2019})}\BibitemShut {NoStop}%
\bibitem [{\citenamefont {Forbes}\ \emph {et~al.}(2021)\citenamefont {Forbes}, \citenamefont {de~Oliveira},\ and\ \citenamefont {Dennis}}]{structuredlight2021}%
  \BibitemOpen
  \bibfield  {author} {\bibinfo {author} {\bibfnamefont {A.}~\bibnamefont {Forbes}}, \bibinfo {author} {\bibfnamefont {M.}~\bibnamefont {de~Oliveira}},\ and\ \bibinfo {author} {\bibfnamefont {M.~R.}\ \bibnamefont {Dennis}},\ }\href {https://doi.org/10.1038/s41566-021-00780-4} {\bibfield  {journal} {\bibinfo  {journal} {Nat. Photonics}\ }\textbf {\bibinfo {volume} {15}},\ \bibinfo {pages} {253} (\bibinfo {year} {2021})}\BibitemShut {NoStop}%
\bibitem [{\citenamefont {Rumelhart}\ \emph {et~al.}(1986)\citenamefont {Rumelhart}, \citenamefont {Hinton},\ and\ \citenamefont {Williams}}]{backpropagation1986}%
  \BibitemOpen
  \bibfield  {author} {\bibinfo {author} {\bibfnamefont {D.}~\bibnamefont {Rumelhart}}, \bibinfo {author} {\bibfnamefont {G.}~\bibnamefont {Hinton}},\ and\ \bibinfo {author} {\bibfnamefont {R.}~\bibnamefont {Williams}},\ }\href {https://doi.org/10.1038/323533a0} {\bibfield  {journal} {\bibinfo  {journal} {Nature}\ }\textbf {\bibinfo {volume} {323}},\ \bibinfo {pages} {533} (\bibinfo {year} {1986})}\BibitemShut {NoStop}%
\bibitem [{\citenamefont {Werbos}(1994)}]{Werbos1994}%
  \BibitemOpen
  \bibfield  {author} {\bibinfo {author} {\bibfnamefont {P.~J.}\ \bibnamefont {Werbos}},\ }\href@noop {} {\emph {\bibinfo {title} {The roots of backpropagation}}}\ (\bibinfo  {publisher} {John Wiley \& Sons},\ \bibinfo {year} {1994})\BibitemShut {NoStop}%
\bibitem [{\citenamefont {Schmidhuber}(2015)}]{nnrev2015}%
  \BibitemOpen
  \bibfield  {author} {\bibinfo {author} {\bibfnamefont {J.}~\bibnamefont {Schmidhuber}},\ }\href {https://doi.org/https://doi.org/10.1016/j.neunet.2014.09.003} {\bibfield  {journal} {\bibinfo  {journal} {Neural Netw.}\ }\textbf {\bibinfo {volume} {61}},\ \bibinfo {pages} {85} (\bibinfo {year} {2015})}\BibitemShut {NoStop}%
\bibitem [{\citenamefont {Sukhov}\ and\ \citenamefont {Dogariu}(2017)}]{Sukhov2017}%
  \BibitemOpen
  \bibfield  {author} {\bibinfo {author} {\bibfnamefont {S.}~\bibnamefont {Sukhov}}\ and\ \bibinfo {author} {\bibfnamefont {A.}~\bibnamefont {Dogariu}},\ }\href {https://doi.org/10.1088/1361-6633/aa834e} {\bibfield  {journal} {\bibinfo  {journal} {Rep. Prog. Phys.}\ }\textbf {\bibinfo {volume} {80}},\ \bibinfo {pages} {112001} (\bibinfo {year} {2017})}\BibitemShut {NoStop}%
\bibitem [{\citenamefont {Jiang}\ \emph {et~al.}(2016)\citenamefont {Jiang}, \citenamefont {Chen}, \citenamefont {Ng},\ and\ \citenamefont {Lin}}]{jiang2016decomposition}%
  \BibitemOpen
  \bibfield  {author} {\bibinfo {author} {\bibfnamefont {Y.}~\bibnamefont {Jiang}}, \bibinfo {author} {\bibfnamefont {J.}~\bibnamefont {Chen}}, \bibinfo {author} {\bibfnamefont {J.}~\bibnamefont {Ng}},\ and\ \bibinfo {author} {\bibfnamefont {Z.}~\bibnamefont {Lin}},\ }\href@noop {} {\bibinfo {title} {Decomposition of optical force into conservative and nonconservative components}} (\bibinfo {year} {2016}),\ \Eprint {https://arxiv.org/abs/1604.05138} {arXiv:1604.05138} \BibitemShut {NoStop}%
\bibitem [{\citenamefont {Yu}\ \emph {et~al.}(2019)\citenamefont {Yu}, \citenamefont {Jiang}, \citenamefont {Chen}, \citenamefont {Liu},\ and\ \citenamefont {Lin}}]{WOS:000486622300012}%
  \BibitemOpen
  \bibfield  {author} {\bibinfo {author} {\bibfnamefont {X.}~\bibnamefont {Yu}}, \bibinfo {author} {\bibfnamefont {Y.}~\bibnamefont {Jiang}}, \bibinfo {author} {\bibfnamefont {H.}~\bibnamefont {Chen}}, \bibinfo {author} {\bibfnamefont {S.}~\bibnamefont {Liu}},\ and\ \bibinfo {author} {\bibfnamefont {Z.}~\bibnamefont {Lin}},\ }\href {https://doi.org/10.1103/PhysRevA.100.033821} {\bibfield  {journal} {\bibinfo  {journal} {Phys. Rev. A}\ }\textbf {\bibinfo {volume} {100}},\ \bibinfo {pages} {033821} (\bibinfo {year} {2019})}\BibitemShut {NoStop}%
\bibitem [{\citenamefont {Zheng}\ \emph {et~al.}(2020)\citenamefont {Zheng}, \citenamefont {Li}, \citenamefont {Jiang}, \citenamefont {Ng}, \citenamefont {Lin},\ and\ \citenamefont {Chen}}]{zheng2020decompchiral}%
  \BibitemOpen
  \bibfield  {author} {\bibinfo {author} {\bibfnamefont {H.}~\bibnamefont {Zheng}}, \bibinfo {author} {\bibfnamefont {X.}~\bibnamefont {Li}}, \bibinfo {author} {\bibfnamefont {Y.}~\bibnamefont {Jiang}}, \bibinfo {author} {\bibfnamefont {J.}~\bibnamefont {Ng}}, \bibinfo {author} {\bibfnamefont {Z.}~\bibnamefont {Lin}},\ and\ \bibinfo {author} {\bibfnamefont {H.}~\bibnamefont {Chen}},\ }\href {https://doi.org/10.1103/PhysRevA.101.053830} {\bibfield  {journal} {\bibinfo  {journal} {Phys. Rev. A}\ }\textbf {\bibinfo {volume} {101}},\ \bibinfo {pages} {053830} (\bibinfo {year} {2020})}\BibitemShut {NoStop}%
\bibitem [{\citenamefont {Zhao}\ \emph {et~al.}(2023)\citenamefont {Zhao}, \citenamefont {Lin}, \citenamefont {Chen}, \citenamefont {Zheng},\ and\ \citenamefont {Ng}}]{zhao2023nanophoton}%
  \BibitemOpen
  \bibfield  {author} {\bibinfo {author} {\bibfnamefont {X.}~\bibnamefont {Zhao}}, \bibinfo {author} {\bibfnamefont {H.}~\bibnamefont {Lin}}, \bibinfo {author} {\bibfnamefont {H.}~\bibnamefont {Chen}}, \bibinfo {author} {\bibfnamefont {H.}~\bibnamefont {Zheng}},\ and\ \bibinfo {author} {\bibfnamefont {J.}~\bibnamefont {Ng}},\ }\href {https://doi.org/doi:10.1515/nanoph-2023-0101} {\bibfield  {journal} {\bibinfo  {journal} {Nanophotonics}\ }\textbf {\bibinfo {volume} {12}},\ \bibinfo {pages} {2019} (\bibinfo {year} {2023})}\BibitemShut {NoStop}%
\bibitem [{\citenamefont {Weiner}(2000)}]{weiner2000femtosecond}%
  \BibitemOpen
  \bibfield  {author} {\bibinfo {author} {\bibfnamefont {A.~M.}\ \bibnamefont {Weiner}},\ }\href {https://doi.org/http://dx.doi.org/10.1063/1.1150614} {\bibfield  {journal} {\bibinfo  {journal} {Rev. Sci. Instrum.}\ }\textbf {\bibinfo {volume} {71}},\ \bibinfo {pages} {1929} (\bibinfo {year} {2000})}\BibitemShut {NoStop}%
\bibitem [{\citenamefont {Zhu}\ and\ \citenamefont {Wang}(2014)}]{zhu2014arbitrary}%
  \BibitemOpen
  \bibfield  {author} {\bibinfo {author} {\bibfnamefont {L.}~\bibnamefont {Zhu}}\ and\ \bibinfo {author} {\bibfnamefont {J.}~\bibnamefont {Wang}},\ }\href {https://doi.org/doi:10.1038/srep07441} {\bibfield  {journal} {\bibinfo  {journal} {Sci. Rep.}\ }\textbf {\bibinfo {volume} {4}},\ \bibinfo {pages} {7441} (\bibinfo {year} {2014})}\BibitemShut {NoStop}%
\bibitem [{\citenamefont {Forbes}\ \emph {et~al.}(2016)\citenamefont {Forbes}, \citenamefont {Dudley},\ and\ \citenamefont {McLaren}}]{forbes2016SLM}%
  \BibitemOpen
  \bibfield  {author} {\bibinfo {author} {\bibfnamefont {A.}~\bibnamefont {Forbes}}, \bibinfo {author} {\bibfnamefont {A.}~\bibnamefont {Dudley}},\ and\ \bibinfo {author} {\bibfnamefont {M.}~\bibnamefont {McLaren}},\ }\href {https://doi.org/http://dx.doi.org/10.1364/AOP.8.000200} {\bibfield  {journal} {\bibinfo  {journal} {Adv. Opt. Photonics}\ }\textbf {\bibinfo {volume} {8}},\ \bibinfo {pages} {200} (\bibinfo {year} {2016})}\BibitemShut {NoStop}%
\bibitem [{\citenamefont {Jiang}\ \emph {et~al.}(2015)\citenamefont {Jiang}, \citenamefont {Chen}, \citenamefont {Chen}, \citenamefont {Ng},\ and\ \citenamefont {Lin}}]{jiang2015universal}%
  \BibitemOpen
  \bibfield  {author} {\bibinfo {author} {\bibfnamefont {Y.}~\bibnamefont {Jiang}}, \bibinfo {author} {\bibfnamefont {H.}~\bibnamefont {Chen}}, \bibinfo {author} {\bibfnamefont {J.}~\bibnamefont {Chen}}, \bibinfo {author} {\bibfnamefont {J.}~\bibnamefont {Ng}},\ and\ \bibinfo {author} {\bibfnamefont {Z.}~\bibnamefont {Lin}},\ }\href@noop {} {\bibinfo {title} {Universal relationships between optical force/torque and orbital versus spin momentum/angular momentum of light}} (\bibinfo {year} {2015}),\ \Eprint {https://arxiv.org/abs/1511.08546} {arXiv:1511.08546} \BibitemShut {NoStop}%
\bibitem [{\citenamefont {Kingma}\ and\ \citenamefont {Ba}(2017)}]{kingma2017adam}%
  \BibitemOpen
  \bibfield  {author} {\bibinfo {author} {\bibfnamefont {D.~P.}\ \bibnamefont {Kingma}}\ and\ \bibinfo {author} {\bibfnamefont {J.}~\bibnamefont {Ba}},\ }\href@noop {} {\bibinfo {title} {Adam: A method for stochastic optimization}} (\bibinfo {year} {2017}),\ \Eprint {https://arxiv.org/abs/1412.6980} {arXiv:1412.6980} \BibitemShut {NoStop}%
\bibitem [{\citenamefont {Paszke}\ \emph {et~al.}(2019)\citenamefont {Paszke}, \citenamefont {Gross}, \citenamefont {Massa} \emph {et~al.}}]{torch2019}%
  \BibitemOpen
  \bibfield  {author} {\bibinfo {author} {\bibfnamefont {A.}~\bibnamefont {Paszke}}, \bibinfo {author} {\bibfnamefont {S.}~\bibnamefont {Gross}}, \bibinfo {author} {\bibfnamefont {F.}~\bibnamefont {Massa}}, \emph {et~al.},\ }in\ \href@noop {} {\emph {\bibinfo {booktitle} {Advances in Neural Information Processing Systems}}},\ Vol.~\bibinfo {volume} {32},\ \bibinfo {editor} {edited by\ \bibinfo {editor} {\bibfnamefont {H.}~\bibnamefont {Wallach}}, \bibinfo {editor} {\bibfnamefont {H.}~\bibnamefont {Larochelle}}, \bibinfo {editor} {\bibfnamefont {A.}~\bibnamefont {Beygelzimer}}, \bibinfo {editor} {\bibfnamefont {F.}~\bibnamefont {d\textquotesingle Alch\'{e}-Buc}}, \bibinfo {editor} {\bibfnamefont {E.}~\bibnamefont {Fox}},\ and\ \bibinfo {editor} {\bibfnamefont {R.}~\bibnamefont {Garnett}}}\ (\bibinfo  {publisher} {Curran Associates, Inc.},\ \bibinfo {year} {2019})\BibitemShut {NoStop}%
\bibitem [{\citenamefont {Virtanen}\ \emph {et~al.}(2020)\citenamefont {Virtanen}, \citenamefont {Gommers}, \citenamefont {Oliphant} \emph {et~al.}}]{2020SciPy-NMeth}%
  \BibitemOpen
  \bibfield  {author} {\bibinfo {author} {\bibfnamefont {P.}~\bibnamefont {Virtanen}}, \bibinfo {author} {\bibfnamefont {R.}~\bibnamefont {Gommers}}, \bibinfo {author} {\bibfnamefont {T.~E.}\ \bibnamefont {Oliphant}}, \emph {et~al.},\ }\href {https://doi.org/10.1038/s41592-019-0686-2} {\bibfield  {journal} {\bibinfo  {journal} {Nat. Methods}\ }\textbf {\bibinfo {volume} {17}},\ \bibinfo {pages} {261} (\bibinfo {year} {2020})}\BibitemShut {NoStop}%
\bibitem [{\citenamefont {Ashkin}\ and\ \citenamefont {Gordon}(1983)}]{opticalearnshaw1983}%
  \BibitemOpen
  \bibfield  {author} {\bibinfo {author} {\bibfnamefont {A.}~\bibnamefont {Ashkin}}\ and\ \bibinfo {author} {\bibfnamefont {J.}~\bibnamefont {Gordon}},\ }\href {https://doi.org/10.1364/OL.8.000511} {\bibfield  {journal} {\bibinfo  {journal} {Opt. Lett.}\ }\textbf {\bibinfo {volume} {8}},\ \bibinfo {pages} {511} (\bibinfo {year} {1983})}\BibitemShut {NoStop}%
\end{thebibliography}%
\bibliographystyle{apsrev4-2}

\end{document}